\documentclass[aps,pre,twocolumn]{revtex4}
\usepackage{amssymb,amsfonts,amsmath,bm}
\usepackage{graphicx}

\usepackage{subfigure}
 \usepackage[T2A]{fontenc}

\usepackage{amssymb, latexsym, amsmath}

 \usepackage{graphicx}

\usepackage{color}

\graphicspath{{art_pic/}{pic/}}
\usepackage{float} 
\usepackage{subfigure}

\begin{document}

\title{Marangoni instability in oblate droplets suspended on a circular frame}

\author{M.A. Shishkin$^{1,2}$, K.S. Kolegov$^{3}$, S.A. Pikin$^{4}$, B.I. Ostrovskii$^{4}$, and E.S. Pikina$^{1,2,5}$. }

\affiliation{$^1$ Landau Institute for Theoretical Physics of the RAS,
142432, Chernogolovka, Moscow region, Russia, \\
$^2$ HSE University, 101000,  Moscow, Russia,  \\
 $^3$ Astrakhan State University named after V.N. Tatishchev,
 414056, Astrakhan,  Russia,\\
$^4$ FSRC "Crystallography and Photonics" of the RAS,
 119333, Moscow, Russia,\\
$^5$ Oil and
Gas Research Institute of the  RAS,  119333, Moscow, Russia}


\begin{abstract}
We study theoretically internal flows in a small oblate droplet suspended on the circular frame.  Marangoni convection arises due to a vertical temperature gradient across the drop and is driven by the surface tension variations at the   free drop interface.   Using the analytical basis for the solutions of Stokes equation in coordinates of oblate spheroid we have derived the linearly independent stationary solutions for Marangoni convection in terms of Stokes stream functions. The numerical simulations of the thermocapillary motion in the drops are used to study the onset of the stationary regime. Both analytical and numerical calculations predict the axially-symmetric circulatory convection motion in the drop, the dynamics of which is determined by the magnitude of the temperature gradient across the drop.
The analytical solutions for the critical temperature distribution and velocity fields are obtained for the large temperature gradients across the oblate drop. These solutions reveal the lateral separation of the critical and stationary motions within the  drops. The critical vortices are localized near the central part of a drop, while the intensive stationary flow is located closer to its butt end.
A crossover to the limit of the plane film is studied within the formalism of the stream functions by reducing the droplet ellipticity ratio to zero value. The initial stationary regime for the strongly oblate drops becomes unstable relative to the many-vortex perturbations in analogy with the plane fluid films  with free boundaries.
\end{abstract}

\maketitle

\section{\label{sec:intro}Introduction}

The Marangoni convection instability is a subject of ongoing research interest at the intersection of fluid mechanics and soft matter physics \cite{Koschmieder1992,vanHook1997,Barash2009,Alexeev2005}. The surface tension of a liquid is usually a decreasing function of temperature. When heat flux is directed across the fluid interface, surface tension gradients due to a temperature variation induce a thermocapillary flow (so called, Marangoni effect). Because of viscosity of the liquid the moving surface gives rise to a shear stress which drives a flow in the film interior. The first manifestation of Marangoni convection goes back to Benard who observed the formation of the characteristic hexagonal convection patterns in flat fluid films subjected to a vertical temperature gradient \cite{Benard}.  Generally, the formation of different cellular flow regimes, including rolls, hexagonal and square patterns, hydrothermal waves, etc. was reported for liquid films of different size and geometry \cite{Koschmieder1974,Davis1987,Levich1962,Gershuni1972,Getling1991,Roisman2015,Nakamura2020,Yoshioka22,Bestehorn2003,Wang2002}. These phenomena are important not only for their influence on the fundamental physics of capillarity, but also for their industrial applications, including chemical engineering, thermal processing of micro-fluidic and electronic devices, evaporation related technology, etc \cite{Kolegov2020,Kolegov2021,Alexeev2005,Roisman2015}. The effects of Marangoni flow become especially noticeable in microgravity conditions where the buoyancy forces are negligible \cite{Orozco,Melnikov2015,Yano2018}.

Here we examine Marangoni flows within a small oblate droplet suspended on the circular frame, Fig. \ref{Figure1}.
Neglecting effects of gravitational force, the  upper and down droplet  surfaces have a shape of  the spherical segments (circular flat lenses), the height of which can vary relative to their lateral dimension
 \cite{Schuring02,Stannarius08,Pikina2020}.
Our theoretical study applies to  simple liquids like  glycerol or silicone oil, polymer melts, as well as the isotropic phase of various complex organic compounds, including liquid crystals. As the droplet  height is less than 1 mm, the gravitational force in Navier-Stokes equation can be neglected, and the convection has a thermocapillary origin \cite{Tam2009,Pikina2022}.

The convection inside a droplet of a curved shape \cite{Tam2009,Barash2009,Sasmal1994,Hu2005,Hu2006,Girard2006,Ristenpart2007,Kita2016}
  appears to be principally different from the conventional Marangoni flows in plane liquid films. A curvature of the drop interface imposes a temperature gradient along its free surface. Thereby,  the tangential thermocapillary force (Marangoni force) always exists at the free drop surface. It causes a fluid flow along its curved interface that is possible for the arbitrarily small vertical temperature gradients. The thermocapillary flow occurs along the free surface of the droplet from the hot area to the cold one, leading to the formation of the torroidal-like vortices within the drop \cite{Pikina2022}, see Fig. \ref{Figure1}.

In this work we present a quantitative description of Marangoni flows within the oblate droplets suspended on the circular frame based on the formalism of the Stokes stream functions \cite{Pikina2022,Happel}. The shape and  axial symmetry of  freely suspended fluid droplet are  well approximated by an oblate spheroid if its height is much less than the median radius. Hence, the oblate spheroid coordinate systems are chosen for the analytical derivations. In our preceding paper \cite{Pikina2022}  we have applied Stokes stream functions method to describe the vortex formation in ellipsoidal isotropic droplets embedded in free standing smectic films. The Stokes stream functions technique is generalized  for the case of the curved fluid interfaces. It is shown that the general solution for the stream functions is represented as a sum over the  basic functions which satisfy the symmetry of the problem and the boundary conditions at the drop interface. Moreover, the original operator method for the solution of differential equations for the stream functions has been developed. The same approach is applied here to obtain the basic set of the stream functions describing the circulatory convection motion in freely suspended drops. The analytical stationary  solutions for the Stokes stream functions as well as the spatial temperature and  velocity distributions for various stages of the convection are obtained. In parallel, the numerical hydrodynamic experiments that model Marangoni flows in oblate droplets suspended on the circular frame are performed. Both the analytical derivations and numerical simulations predict the axially-symmetric circulatory convection motion within the droplet determined by the Marangoni effect at the droplet free surface.

In general, the convection patterns have a  shape of individual torroidal-like vortices. The analysis of the influence of the sticking area along the drop equator on the character of thermocapillary motion is performed.
It is shown that the fastest fluid flow along the vortex trajectory occurs not at the end face of a drop as in the case of a fully free drop, but at a certain distance from it.
 However, the shape and dynamics of vortices in the drop interior are almost  not affected.
Thereafter, the critical regime of Marangoni convection is analyzed for the large temperature gradients across the drop. It is found that the critical and stationary solutions are lateraly separated within the flattened drops. At that the critical vortices are localized within the central part of a drop,  while the intensive stationary flow is located closer to its butt end.
  Finally,   a crossover to the limit of the flat fluid film has been investigated. Such a crossover is made within the formalism of the stream functions by reducing the droplet ellipticity ratio to zero value. We have shown that for such flattened droplets and under action of the considerable temperature gradients there is a tendency for the formation of a series of vortices distributed within the plane of the drop.  The stability analysis  reveals  that the initial stationary regime for the strongly oblate drops becomes unstable relative to  many-vortex perturbations in analogy with the plane fluid films.

\section{Stationary convection}

\subsection{Statement of the problem. Governing equations and boundary conditions}

\begin{figure}
    \includegraphics[width=0.85\linewidth]{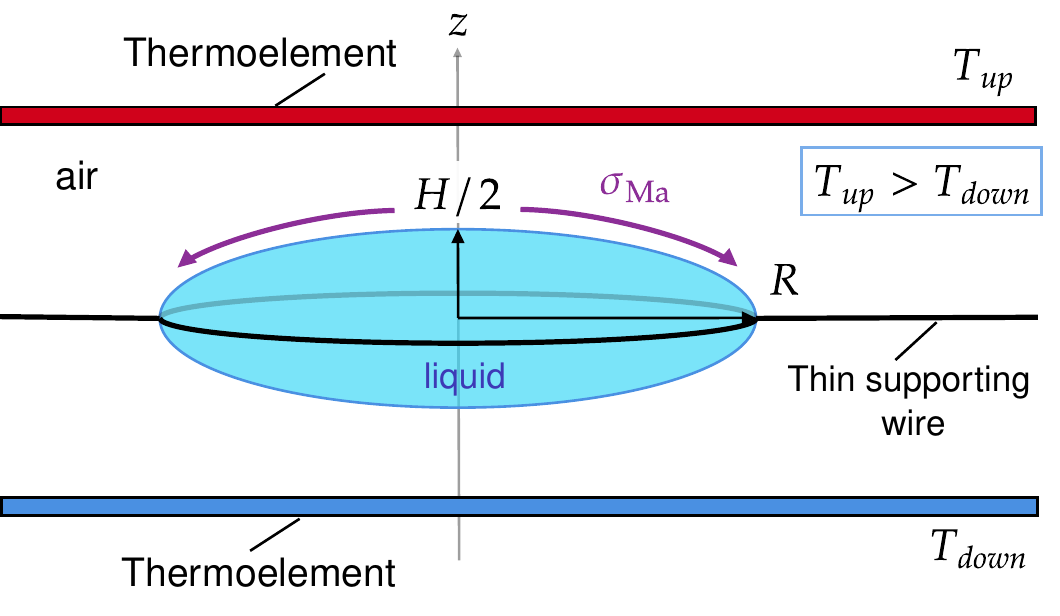}
        \caption{Schematic view of a fluid droplet suspended on a supporting circular wire. The drop has a lenslike shape and is symmetric relative to the horizontal plane. The drop is approximated by the oblate spheroid with a  large, $R$, and small, $H/2$, semiaxes, where $R$ and $H$ are the drop base radius and height, respectively. $\sigma_{\mathrm{Ma}}$ defines  Marangoni tension acting along the drop free surface in the direction as indicated; $T_{up}> T_{down}$.}
    \label{Figure1}
\end{figure}

 The geometry of our problem is shown in Fig. \ref{Figure1}. The central circular crossection of a drop in the form of an oblate spheroid is parallel to the $x-y$ plane. The z axis corresponds to the axis of symmetry of the droplet, and the origin of the coordinate frame is taken in the center of the drop. Two thermoelectric elements above and below the drop are used to create the vertical temperature gradient across it. Due to a symmetry of the problem, the thermocapillary flow within the fluid drop is possible for both directions of the thermal gradient across it: from bottom to top and inverse. For further derivations the direction of the heat transfer from the hot upper plate to the cold bottom plate is chosen.
 It corresponds to the positive direction of the temperature gradient $\partial T/\partial z$
  $\,(T_{dn} < T_{up})$,  Fig. \ref{Figure1}, and  ensures the absence of the Rayleigh convection in the surrounding air.

The geometry of the fluid drop suspended on the circular frame is determined by the given  drop volume and the frame properties.
As is shown in our preceding paper \cite{Pikina2022}, the drop shape is well approximated by an oblate spheroid under condition that its half height ($H/2$) is much less than the drop base radius $R $, $H\ll\,2\,R\,$.
 Following \cite{Pikina2022}, we use below the oblate spheroid orthogonal coordinates $u, \xi,  \varphi\,$
        related  to the rectangular  Cartesian coordinates $(x,y,z)$ by the  vector representation \cite{Happel,NLebedev65,NLebedev652}:
\begin{eqnarray}
  \left(\begin{array}{c}
 x \\ y \\ z
 \end{array} \right)
\, =
\,  \left(\begin{array}{c}
 \,c\,\sqrt{1+\xi^2}\,\sqrt{1-u^2}\,\cos[\varphi]\\
 c\,\sqrt{1+\xi^2}\,\sqrt{1-u^2}\,\sin[\varphi] \\
 c\,u\,\xi\,
 \end{array} \right) \  \
 \label{1elc}
 \end{eqnarray}
where
\begin{eqnarray}
 \ - 1\, \le u \le \, 1 \, , \
0\, \le \xi < \,\infty\, ,
 0\, < \varphi \, \le\,  \,2\,\pi \,\, . \qquad  \  \  \label{1elc1}
\end{eqnarray}
The corresponding metric coefficients can be written as
    \begin{eqnarray}
    h_u = c \sqrt{\frac{\xi^2 + u^2}{1-u^2}} \ ,
    h_\xi = c \sqrt{\frac{\xi^2 + u^2}{1+\xi^2}} \ , \
       \nonumber \\
    h_\varphi = c\sqrt{1+\xi^2}\sqrt{1-u^2}
      \  . \ \label{aLamexi}
      \end{eqnarray}
      In above notations the surface of an oblate spheroid is determined by an equation $\xi =\xi_0$, see Appendix \ref{sec:A}.
Using the ellipsoidal coordinates an oblate spheroid  can be characterised
 by the semiaxes ratio
 $H/(2 R)\ll\,1$, where $c\,\xi_0\equiv H/2$ and $H\,\sqrt{1+\xi_0^2}/(2 \xi_0)\equiv\,R $ are the small and  large semiaxis of ellipsoid, respectively, $c=R\,\big(1-H^2/(4 R^2)\big)^{-1/2}$ is a focus distance of ellipsoid. The  parameter  $\xi_0=\frac{H}{(2 R)}\,\big(1-H^2/(4 R^2)\big)^{-1/2}$,  $\xi_0\ll 1$, determines
the ellipticity ratio of the drop.

       The hydrodynamic flows in the droplet are described by Navier-Stokes and the continuity equations for the incompressible fluid  as well as the heat transfer equation \cite{Gershuni1972,Getling1991,vanHook1997,Landau6,Falkovich,Pikina2022}. Because the height of the droplets is pretty small (less than 1 mm), the gravitational force in Navier-Stokes equation can be neglected and the convection has a thermocapillary origin \cite{Tam2009,Pikina2022}. Marangoni convection in the drop is considered as  small perturbations of the  temperature and velocity fields driven by the surface tension variations at the drop free interfaces. As in any normal fluid the surface tension, $\gamma$, is a decreasing function of temperature
\begin{equation}
   \gamma = \gamma_0 -  \varsigma \,T'
      \ , \
      \label{gam}
  \end{equation}
where $T' =  (T_{dr} -  \bar{T})$, $T_{dr}$ is a current drop temperature, $\bar{T}$ is some  constant temperature (far from the drop at $z=0$), and $ \varsigma\,$  is the tension temperature coefficient ($ \varsigma\,>0$).
 Below we omit the symbol $'$ for simplification
of  further derivations.

In that follows we calculate the stationary thermocapillary flows and vortex formation in  freely suspended drops for  relatively small temperature gradients  in a framework of the conventional linear perturbation theory.
The main assumptions  that we use are identical to  Boussinesq approximation (see Appendices \ref{sec:A} and \ref{sec:C}, compare also with \cite{Gershuni1972,Getling1991,vanHook1997,Landau6}).

It is convenient to solve  hydrodynamic equations of Marangoni convection for the axially symmetric drops in terms of 2D Stokes stream functions $\psi[u_, \xi]$   \cite{Happel,psi}.  In accordance with \cite{Happel}, the velocity field is related to  the  stream function by the following equation written in oblate spheroidal coordinates
  \begin{eqnarray}
      \mathbf{v} = \frac{1}{h_\varphi}\,[{\bf e}_{\varphi} \,\times\, \nabla \psi] \ . \
       \label{vsp}
              \end{eqnarray}
 where ${\bf e}_{\varphi}$ is the  azimuthal unit vector  oriented beyond the page (sheet) plane, (see Appendix A).
    After substitution of Eq. (\ref{aLamexi}) to Eq. (\ref{vsp}) one  obtains
       \begin{eqnarray}
      \mathbf{v} =  \, - \frac{{\bf e}_{u}}{h_\xi h_\varphi}\partial_\xi \psi \, +\, \frac{{\bf e}_{\xi}}{h_u h_\varphi}\partial_u \psi   \ , \
       \label{bfu}
              \end{eqnarray}
              where ${\bf e}_{\xi}$ and ${\bf e}_{u}$  are the unit vectors along $\xi$ and $u$ axis, respectively, see Appendix A.

Next, to obtain the dynamic equation for the Stokes stream function it is convenient to introduce the vorticity, $\vec{\mathbf{\varpi}}$, \cite{Happel}
               \begin{eqnarray}
     \vec{\mathbf{\varpi}} = \,[\nabla\,\times\,  \mathbf{v}] \ ,\ \
       \label{vort1}
              \end{eqnarray}
              for which the continuity equation is satisfied  automatically.
Applying the standard rules of the vector differentiation to the axial
vorticity vector, one can express  the resulting equation through the single variable $\psi$ \cite{Happel}. While doing it, we replace the standard Navier-Stokes equation for the stationary flow regime in the linear approximation over ${\bf v}$
with the following equation for the stream function
 \begin{eqnarray}
\hat{\text{E}}^2 \,\big( \hat{\text{E}}^2\,\psi \big)
 \,=\,0
    \ , \
    \label{big1}
                      \end{eqnarray}
where operator $\hat{\text{E}}^2$ has a form
\begin{eqnarray}
 \hat{\text{E}}^2 \psi =   \frac{1}{c^2\,\big(u^2  + \xi^2 \big)}\,
\Big\{ \,(1  + \xi^2)\,\frac{\partial^2 \psi}{\partial\xi^2}
 +   (1 - u^2)\,\frac{\partial^2 \psi}{\partial u^2}\, \Big\}
     \, . \  \ \ \label{StrF2}
                  \end{eqnarray}
The Eq. (\ref{big1})  is accompanied by the stationary equations for the temperature distributions across the drop and in the surrounding air  \cite{Pikina2022}
\begin{eqnarray}
   \chi\,\Delta T \ \,=\, \,({\bf v}\,\mathbf{\nabla})\,{T}
        , \    \label{big2}  \\
       \Delta T_{a} \,=\, 0 \
        , \    \label{big2a}
\end{eqnarray}
where $\chi$ is a coefficient of temperature conductivity, $\chi = \varkappa \,(\rho_0 c_p)^{-1}$, ($\varkappa\,$ is the thermal conductivity, $\rho_0$ and $c_p$ are  fluid density  and specific heat, respectively).

Let us now formulate  the boundary conditions for the oblate spheroidal drop suspended on a frame. At the surface of a drop ($\xi=\xi_0$) the boundary condition for the normal fluid velocity component reads
                  \begin{eqnarray}
 v_{\mathbf{n}} =  v_{\xi} = 0 \ \ ({\hbox{at}} \, \xi=\xi_0 ) \ , \
\label{vxibc}
                  \end{eqnarray}
                that is the condition  of the absence of the material flow  through the boundary  drop surface. At
           the same time in the sticking area along the drop equator we have  for the tangential velocity component
                  \begin{eqnarray}
v_{\tau} = -\,  v_{u} = 0     \quad \ ({\hbox{at}} \, \xi=\xi_0\, , \, u
      \in [ -u_s, u_s])
             \, , \
             \label{xiubc0}
                       \end{eqnarray}
                          where the interval  $[ -u_s, u_s]$
               determines an extension of the boundary with the sticking (no-slip) conditions (in the contact with a ring)  along the drop interface.

   Another type of the boundary conditions at the  drop surface  corresponds to a stress balance both in the normal and tangential directions. The  boundary condition  for the balance of tangential viscous  and  Marangoni   forces  on  the  free   boundary  of the ellipsoidal drop  is given by  the expressions \cite{Pikina2022}
\begin{eqnarray}
&&\sigma_{u\xi}  \,=\eta \Big[ \frac{\partial_u v_\xi}{h_u}+
    \frac{\partial_\xi v_u}{h_\xi} - \frac{v_u}{h_\xi}\frac{\xi}{\xi^2+u^2}
    \Big] \
    \nonumber \\
  && =  \,  \frac{\partial_u \gamma}{h_u} \quad ({\hbox{at}} \, \xi=\xi_0, \, u
        \in   [{\hbox{free surface region}}]  )
             \,  , \ \    \label{xiubc}
                  \end{eqnarray}
                  where $\eta\,$ is a dynamical  viscosity coefficient of liquid.
     As for the normal stress balance,
      it is almost satisfied due to the fact that the oblate shape of a drop does not actually change in the process of convection. One can check that the pressure deviation induced by the temperature difference across the drop boundary is negligibly small:  $\delta R/R \sim \delta \gamma /\gamma \ll 1$ (i.e. due to small surface tension $\gamma$ variations).

     Next we consider the boundary conditions for the temperature deviations and the heat fluxes
                 \begin{eqnarray}
  T_{a }\big\vert_{\xi=\xi_0}  = \, T \big\vert_{\xi=\xi_0}
             \, , \ \label{bcT0}
             \\
          T_{a}\big\vert_{\xi\to\infty}= \, C_{air}\,z\,  = \, C_{air}\,{c\,u\,\xi} \,
           \, , \ \label{bcT00}
      \end{eqnarray}
              \begin{eqnarray}
     \varkappa\,  \frac{\partial T}{\partial {\xi}}\Big\vert_{\xi=\xi_0} =
     \varkappa_{air}\,\frac{\partial {{T}_{a }}}{\partial {\xi}}\Big\vert_{\xi=\xi_0}
                 \, , \ \label{bcT2f}
      \end{eqnarray}
      which determine the boundary conditions of the equality of the temperature  deviations and the normal heat flux at the air-drop interface, where $C_{air}$ is a uniform temperature gradient across the air.

\begin{figure*}
\subfigure[]{
\includegraphics[scale=0.8]{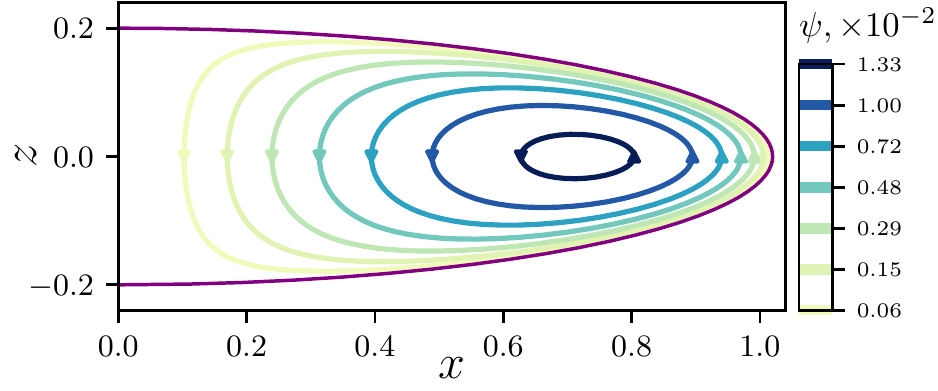}\label{Figure3a}}
\subfigure[]{
\includegraphics[scale=0.8]{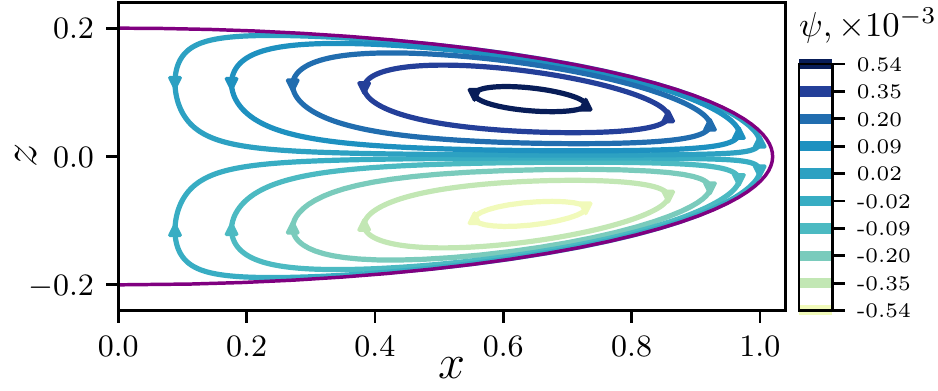} \label{Figure3b}}
\subfigure[]{
\includegraphics[scale=0.8]{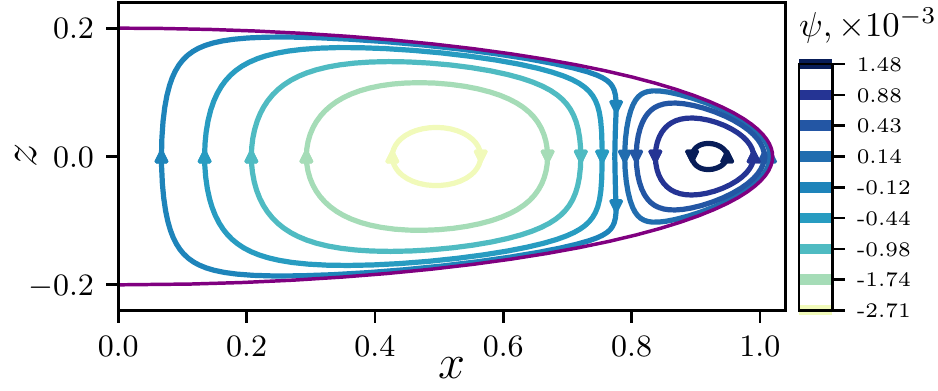}\label{Figure3c}}
\subfigure[]{
\includegraphics[scale=0.8]{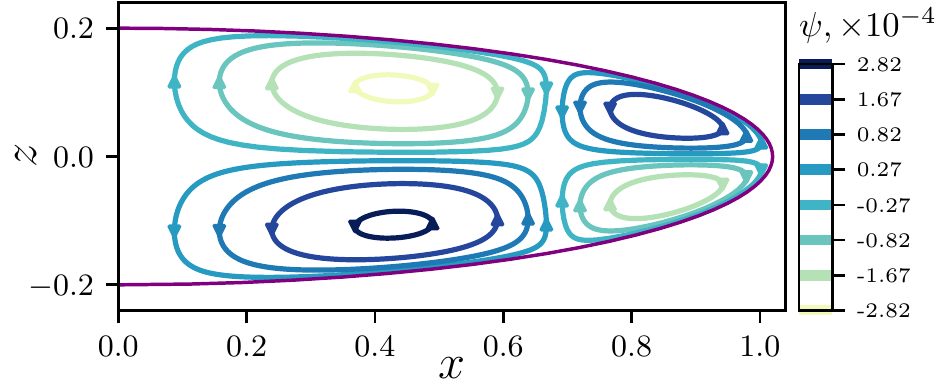} \label{Figure3d}}
\caption{Streamlines corresponding to the basic stream functions  $\psi_{3}$, $\psi_{4}$, $\psi_{5}$ and $\psi_{6}$, from \subref{Figure3a} to \subref{Figure3d}, respectively; $\xi_0 = 0.2$; the $\xi_0$ value determines the ellipticity ratio of the droplet and is equal to  $\xi_0 = H/(2 c)=\frac{H}{(2 R)}\,\big(1-H^2/(4 R^2)\big)^{-1/2}$, (see section II A).
 All lengths  are shown in the dimensionless form, being scaled by
 $c = R\,\big(1-H^2/(4 R^2)\big)^{-1/2}$.}
\label{Figure3}
\end{figure*}


  \subsection{Stokes stream functions within  the oblate spheroid drop}

The general solution of Eq.  (\ref{big1}) for the stream function $\psi$ represents the sum over the basic functions $\psi_n[u, \xi]$, which satisfy the symmetry of the problem and the boundary conditions at the drop interface \cite{Pikina2022}. To derive the expressions for various $\psi_n$ we introduce a set of operators $ \hat{\mathcal{F}}$ and $\hat{\mathcal{X}}$, that
simplify the calculations of the stream functions, where
\begin{eqnarray}
    \hat{\mathcal{F}} \, = \, (1-u^2)\, \partial_u^2 \ , \
    \label{F}
    \end{eqnarray}
     with eigenfunctions $\mathcal{F}_n $
\begin{align}
  \hskip-.3true cm       \hat{\mathcal{F}}\mathcal{F}_n =  -n(n+1) \mathcal{F}_n \, , \  \mathcal{F}_n[u] = \frac{P_{n+1}[u] - P_{n-1}[u]}{2n+1} \, , \,
    \label{Fn}
\end{align}
where $P_n[u]$ are Legendre polynomials  of the order $n$. In turn,
 \begin{eqnarray}
    \hat{\mathcal{X}} \, = \, (1+\xi^2) \, \partial_\xi^2 \ ,  \
    \label{opXi}
\end{eqnarray}
with eigenfunctions $\mathcal{X}_n[\xi]$
\begin{align}
       \hat{\mathcal{X}} \mathcal{X}_n = n(n+1)\mathcal{X}_n\,,\; \mathcal{X}_n = \frac{\Xi_{n+1}[\xi]-\Xi_{n-1}[\xi]}{2n+1} \, , \
    \label{Xin}
\end{align}
where the functions $\Xi_n[\xi]$  are obtained by transformation to  real presentation  by redefinition Legendre polynomials of imaginary argument:
\begin{equation}
\Xi_n[\xi]= \begin{cases}
P_{n}[i\xi] \text{ for even n},\\
(-i)\cdot P_{n}[i\xi ] \text{ else} \ . \
 \end{cases}
     \label{rexi}
\end{equation}
  The operator $\hat{E}^2$, introduced in  Eqs. (\ref{big1}) -- (\ref{StrF2}),
 can be rewritten in the oblate spheroid coordinates as
\begin{equation}
    \hat{E}^2 = \frac{\hat{\mathcal{X}} + \hat{\mathcal{F}}}{c^2(\xi^2+u^2)} \ . \
\end{equation}
The above equation indicates that the kernel of the operator $\hat{E}^2$ is  $\{\mathcal{X}_n \mathcal{F}_n\}$, see Eqs. (\ref{Fn}), (\ref{Xin}).
 It  means that the general smooth solution of Navier-Stokes equation (\ref{big1})  for the stream function $\psi\,$ can be written in terms of the eigenfunctions of operators $\hat{\mathcal{F}} $ and $\hat{\mathcal{X}}$. In accordance with \cite{Pikina2022}, the solution for the stream function $\psi\,$  is represented as a series
\begin{equation}
    \psi = \sum_{n>2} c_n \, \big(\mathcal{F}_n \,\mathcal{X}_{n-2}  +
    \mathcal{F}_{n-2}\, \mathcal{X}_{n} \big) + \sum_{n\ge1} c_{n o}\, \mathcal{X}_n\,\mathcal{F}_n \ .  \
    \label{ser}
\end{equation}
The boundary condition of Eq. (\ref{vxibc}) with account to Eq. (\ref{bfu}) takes the form:
  \begin{eqnarray}
\psi[\xi_0, u]  \,=\,0 \
    \, . \  \ \ \label{BCS1}
                  \end{eqnarray}
The condition (\ref{BCS1}), together with the linear independence of the functions $\mathcal{F}_n$,  restricts the possible set of the constants $c_{n o}$.
Finally, the full analytical solution of Eq. (\ref{big1}) has the form
\begin{equation}
\psi[\xi, u]\, = \,\sum_{n>2}c_n \psi_n[\xi, u]\, \ , \
    \label{fulsol}
\end{equation}
 where
 the  general expression for the $n-$th basic stream function $\psi_n$
with account to the boundary condition, Eq. (\ref{BCS1}) can be written as  \cite{Pikina2022}
\begin{eqnarray}
\psi_n[\xi, u]\, = \, \mathcal{F}_n[u] \Big( \mathcal{X}_{n-2}[\xi] \,- \, \frac{\mathcal{X}_{n-2}[\xi_0]}{\mathcal{X}_{n}[\xi_0]}\, \mathcal{X}_{n}[\xi] \Big)
\nonumber \\
 + \, \,\mathcal{F}_{n-2}[u] \Big( \mathcal{X}_{n}[\xi]
\,- \, \frac{\mathcal{X}_{n}[\xi_0]}{\mathcal{X}_{n-2}[\xi_0]}\,\mathcal{X}_{n-2}[\xi] \Big) \  . \ \ \
 \label{GenS}
  \end{eqnarray}
 The number of terms in expansion (\ref{fulsol}) is determined by  the number, $N_r$, of  the basic functions needed for the convergence of this  expansion (see further).
  The examples of  series of the basic functions of the lowest order (from $n=3$ to $n=6$)  are shown in Fig. \ref{Figure3}.
  The number of vortices along the  long drop semiaxis $a$  for each $\psi_n$ increases with  $n$.
The constants $c_n$  in  Eq. (\ref{fulsol}) can be found using
the balance condition for the tangential forces, Eq. (\ref{xiubc}),
 and sticking condition
 at the drop equator for the drop suspended on the circular ring, Eq. (\ref{xiubc0}).

\subsection{Results and discussion}

In this section we present the stationary solution of Eq. (\ref{big1}) in terms of a stream function $\psi$ for  an oblate spheroidal drop suspended on a solid ring. Prior to considering this problem, it is worth analyzing   the minimal model of Marangoni convection in the ellipsoidal drop with a fully free surface, without any suspension loop. Despite of the ideal character of this model, it has a clear advantage of providing the  exact analytical solution for hydrodynamic flows in a drop, and allows the limiting cases of a spherical drop and a plane fluid film with  free boundaries to be investigated. The obtained solutions for the free drop will be used further to describe the influence of the sticking conditions along the drop equator on the character of  the thermocapillary flow within it. The later results can be directly implemented to the experimentally accessible drop geometry with a suspension ring.  The model of an oblate drop with a fully free surface will be applied also for the analysis of the stability of the stationary Marangoni solutions relative to the increase of the temperature gradient across the drop.

\subsubsection{ Analitical results for the fully free drop}

We start with the governing equations of Marangoni convection describing the thermal energy transport inside the drop and the temperature distribution in the surrounding air, Eqs.  (\ref{big2}), (\ref{big2a}).
We are using below a conventional linear perturbation theory in a frame of which the temperature distributions in the drop and in  the surrounding air are represented as: $T = T_0 + T_1$ and $T_a = T_{0\, air} + T_{1\, air}$.
 For the relatively small Marangoni numbers, Ma $\,\ll\,(R/H)^2$,  one can disregard the heat exchange in  the convection process that leads to  inequalities $T_1\ll T_0$ and $T_{1\, air}\ll T_{0\, air}$ (see Appendices \ref{sec:A}, \ref{sec:B},  and inequalities (\ref{uneqvT4}), (\ref{uneqv4}) and (\ref{uneqvt}) there). As a result,
      Eqs.  (\ref{big2}), (\ref{big2a}) can be simplified to Laplace equations for $T_0$ and $T_{0\, air}$:
\begin{eqnarray}
         \Delta T_0 = 0 \ , \
    \label{T01} \\
     \Delta T_{0\, air} = 0 \ , \
    \label{T0air}
\end{eqnarray}
 Using Eqs. (\ref{fulsol})--(\ref{T0air}), the boundary conditions, Eqs. (\ref{vxibc})--(\ref{bcT2f}), and intermediate calculations, (\ref{solT1})--(\ref{lapl2b}), presented in Appendix \ref{sec:B}, the temperature distribution within the free drops is obtained.

In that follows we present all  variables in  the dimensionless form, using the appropriate scaling relations. The lengths are scaled by
$c=R\,\big(1-H^2/(4 R^2)\big)^{1/2}$, the stream function by $c^2\chi/H$,  velocities by $\chi/H$,  the temperature by $AH$ and time by $c\,H/\chi$.
According to Eq. (\ref{solTd0}), $T \approx T_0 = A z$, where $A$ is a uniform
temperature gradient across the drop. The same temperature distribution in the  dimensionless form reads: ${T}\approx  {T}_0= u\xi/(2\xi_0)$.

\begin{figure}[h]

\includegraphics[scale=0.7]{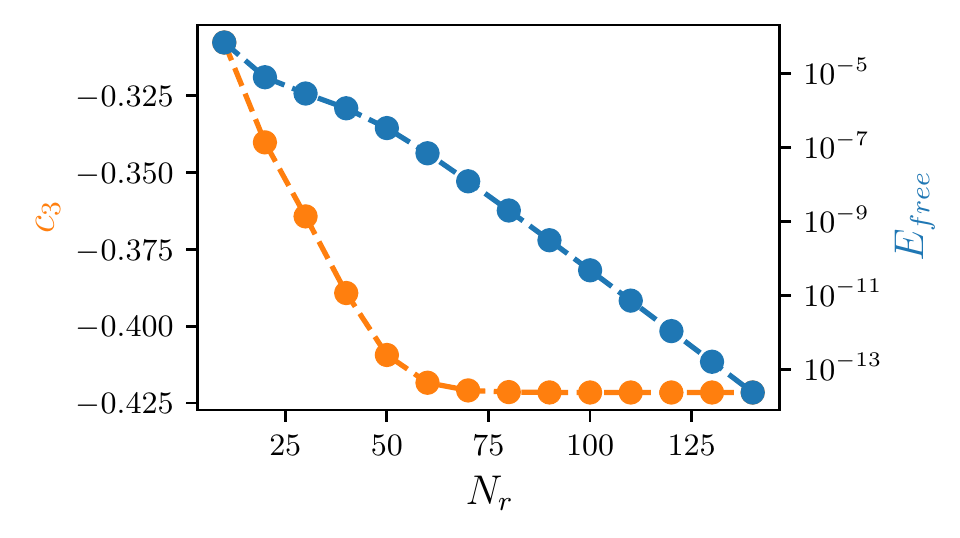}
\caption{Illustration of  a steep decay of the mean square error in Marangoni condition and corresponding convergence of the first coefficient $c_3$ in expansion, Eq. (\ref{fulsol}), for the stream function  with increase of the number, $N_r$, of terms in the expansion ($\xi_0=0.1, \, \mathrm{Ma}=1$).}
\label{c3fr}
\end{figure}

\begin{figure}
\subfigure[]{
\hskip.5true cm \includegraphics[scale=0.78]{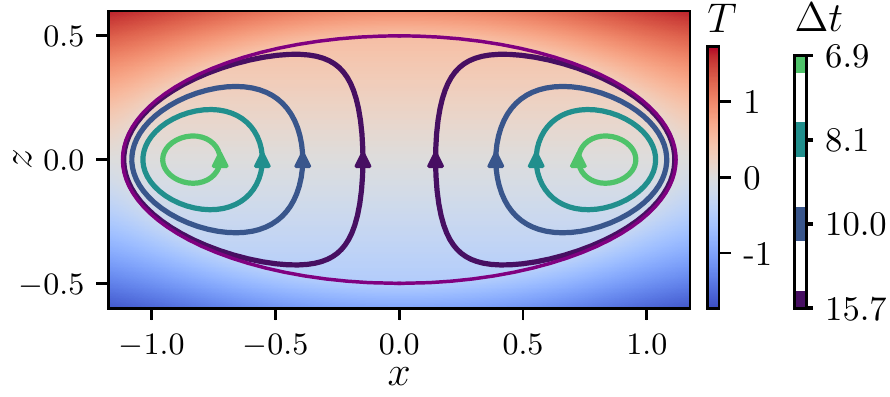}\label{Figure5a}}
\subfigure[]{
\includegraphics[scale=0.8]{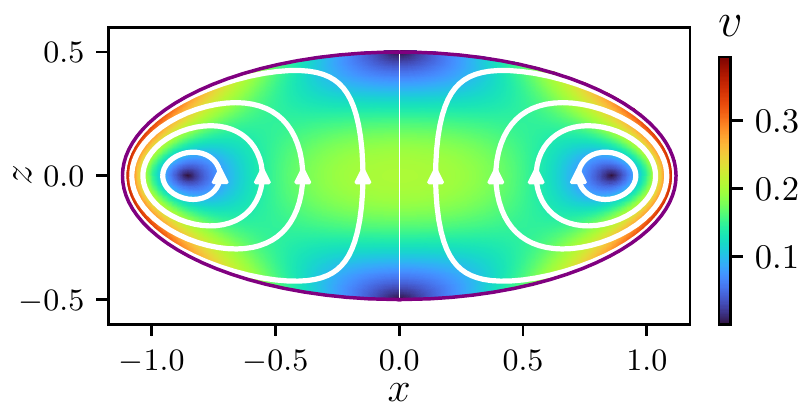} \label{Figure5b}}
\caption{Example of the convective motion within the drop with a fully free surface (Ma $=1,\,\xi_0=0.5, \,  \kappa=0.2$, where $\kappa = \varkappa_{air}/\varkappa$ is a relative heat conductivity).
The $\xi_0$ value determines the ellipticity ratio of the droplet and is equal to $\frac{H}{(2 R)}\,\big(1-H^2/(4 R^2)\big)^{-1/2}$.  All the lengths, temperature $T$, circulation period  $\Delta t$ (a) and velocity modulus $v$ (b) are shown in the dimensionless form.
The dimensional values of velocity $v$ and circulation period $\Delta t$  can be deduced  from the  dimensionless ones through  multiplication by the corresponding scaling parameters: velocity by   $(\chi/H)\hbox{Ma} =  \varsigma A H /\eta$ and period $\Delta t$ by $ (1/\hbox{Ma})\,c/((\chi/H)) = {\eta}/{(2\xi_0\, \varsigma A)}$.
 }
\label{Figure5}
\end{figure}

The Marangoni boundary condition (\ref{xiubc}) in terms of the dimensionless variables can be written as:
\begin{eqnarray}
&&{\left(\partial_\xi-\frac{(\xi_0^2+u^2)}{2\xi_0}\partial_\xi^2 \right){\psi}}\Big\vert_{\xi=\xi_0}
\nonumber \\
&&    =\,-\,\frac{\mathrm{Ma}}{2\xi_0 \sqrt{1+\xi_0^2}} (\xi_0^2+u^2)^{3/2}\, (1-u^2)\, \partial_u T_0[u, \xi_0]\, , \qquad
\label{BcMa}
\end{eqnarray}
where
\begin{equation}
   \mathrm{Ma} = \frac{\varsigma H^2 A}{\chi \eta} \
\label{Ma}
\end{equation}
is Marangoni number.
The substitution of the expression for $T_0$ and an expansion for the stream function $\psi$ to Eq. (\ref{BcMa}) yields the equation
\begin{equation}
    \sum_{j>2} c_j l_j[u] \,=\, r[u]\, , \
    \label{BCsim}
\end{equation}
where
\begin{eqnarray}
    l_j[u] = {\left(\partial_\xi-\frac{(\xi_0^2+u^2)}{2\xi_0}\partial_\xi^2 \right){\psi_j}}\Big\vert_{\xi=\xi_0} \, , \
\label{lj}
\end{eqnarray}

\begin{equation}
r[u] =  \,-\,\frac{\mathrm{Ma}}{4\xi_0\sqrt{1+\xi_0^2}}(\xi_0^2+u^2)^{3/2}\, \,(1-u^2) \, . \
\label{ru}
\end{equation}

Unfortunately, as it was shown in our previous paper  \cite{Pikina2022}, there is no way to  solve  Eq. (\ref{BCsim}) and  derive all  coefficients $\{c_j\}$ exactly due to the irrationality in $r[u]$.
 To overcome this problem  the  solution of Eq. (\ref{big1}) is presented as a sum over the limited  number, $N_r$, of the $j-$th basic stream functions  $\{\psi_j[\xi, u] \}$: ${\psi^{(N_r)} = \sum_{j>2}^{N_r} c_j \psi_j}$.
  The optimal set of coefficients $\{c_j\}_{opt}$ should satisfy the following convergence criterion: the norm of deviation of Eq. (\ref{BCsim}) from zero, $E^{free}[\{c_j\}_{opt}]$, reaches  the minimum mean square value for the optimal $N_r$-measured set of $\{c_j\}_{opt}$ (compare with  \cite{Pikina2022})
\begin{equation}
    E^{free}[\{c_j\}] = \int_{free} \left(\sum_{j>2}^{N_r} c_j l_j[u] - r[u]\right)^2 \frac{d{u}}{1-u^2} \ ,
    \label{BCfr}
\end{equation}
where the multiplier  $1/(1-u^2)$  is placed to  account for the orthogonality of the functions $\{\mathcal{F}_j\}$.
The functional $E^{free}[\{c_j\}]$ should converge to zero in the limit
 $N_r\to\infty$.
  In  Fig.   \ref{c3fr} one can  see the fast exponential decay of the function $E^{free}[\{c_j\}_{opt}]$, and  a simultaneous stabilization of one of the coefficients, $c_3$, in the expansion (\ref{BCsim}) for $N_r >75$. The same is true for the  other terms in  the expansion (\ref{BCsim}).

	\begin{table} \caption{ The parameters   used for the numerical calculations} \centering
		\begin{tabular}{|p{0.13\linewidth}|p{0.5\linewidth}|p{0.31\linewidth}|} \hline
			Symbol& Parameter& Value [unit of measurement]\\ \hline
			$\varkappa $& thermal conductivity of liquid &0.12 [W/(mK)]\\
			$\varkappa_a$& thermal conductivity of air& 0.026 [W/(mK)]\\
			$c_p$& specific heat capacity of liquid& 2500 [J/(kg~K)]\\
			$c_a$& specific heat capacity of air& 1000 [J/(kg~K)]\\
			$\eta$& dynamic viscosity& $1.4 \times 10^{-2}$ [s~Pa]\\
			$\rho $& liquid density& 1200 [kg/m$^3$]\\
			$\rho_a$& air density& 1.2 [kg/m$^3$]\\
			$\varsigma$& tension  temperature  coefficient &$5\times 10^{-5}$ [N/(m~K)]\\
			$H$& droplet height &20 [$\mu$m]\\
						$R$& droplet radius &100 [$\mu$m]\\
			$T_\mathrm{dn}$& bottom plate temperature &324 [K]\\
			$T_\mathrm{up}$& top plate temperature &334 [K]\\ \hline
		\end{tabular} \label{tab:Parameters}
	\end{table}

    Therefore, it confirms the regular convergence provided by our procedure and  the correctness of our approach within the derivation of the general solution for the stream function.
  	The example of the calculated  vortex motion within a  fully free drop is shown in Fig. \ref{Figure5}. All the variables are given in the dimensionless form. While doing calculations we use  the material  and transport parameters of the typical liquids, see Table I.
   The height $H$ of the droplets can be changed relative to their diameter by varying  the ellipticity ratio $\xi_0$.
   To characterize the circulating Marangoni flows within the drops we use the time period, $\Delta t$, i.e., the time interval required for the movement along the closed stream line. The time period is  defined as
      \begin{equation}
\Delta t = \, \oint d t \,= \, \oint \frac{d l}{v} \ , \
\label{Dt}
\end{equation}
where $v$  is a velocity modulus along the trajectory of the flow and $dl$ is a tangential element of the curved trajectory. The magnitude of the circulation period $\Delta t$ for various closed vortex lines can be directly measured in experiment  by tracing the circulatory movement of the properly selected micro-particles within the drop.
In  Figs. \ref{Figure5}, \ref{Figure8}, \ref{Figure10} the examples of the stationary convection motion in oblate drops of various flatness are shown as calculated for Marangoni number Ma =1. Because of a linear response of a system on variation of Ma to obtain the dimensional values of  the velocity $v$ and circulation period $\Delta t$ for an arbitrary Ma
one should multiply their  dimensionless values by the
corresponding  scaling parameters. For the velocity $v$ this parameter is
$\frac{\chi}{H}\,\hbox{Ma} =  \frac{\varsigma \,A \,H }{\eta}\,$,
 while for the  period $\Delta t$ this  is
$\frac{ c H}{\chi\,\hbox{Ma}}  \,=  \frac{\eta}{2\,\xi_0\, \varsigma\, A}\,$.

In Figs. \ref{Figure8}, \ref{Figure10} the circulating flows within a fully free drop are shown in comparison with the vortices calculated for the droplets suspended on the solid ring.
  In Figs. \ref{Figure5}, \ref{Figure6a}, \ref{Figure6b}, \ref{Figure10a} one can see that the fastest  circulatory  motion  in ellipsoidal drops with a fully free surface occurs near their end faces.
This observation is not surprising. Since
  the tangential temperature gradient is parallel to the drop surface, Marangoni force attains the largest values near the end faces of the  drops, where the corresponding temperature gradient is infinite.

The magnitudes of the fluid velocity and circulation period $\Delta t$ depend on  Marangoni number, Ma, the drop ellipticity ratio $\xi_0$ and the drop radius $R$. If the geometrical parameters of a drop are fixed, the flow velocity increases upon the Ma increase, while the corresponding time period $\Delta t$ diminishes.
For a given value of Ma (i.e. of the temperature gradient across the drop) the  flattened drops, which are characterized by a smaller ellipticity ratio $\xi_0$, show a lesser velocity magnitude $v$. Accordingly, their circulation period $\Delta t$ increases, see Figs. \ref{Figure5}, \ref{Figure8}, \ref{Figure10}.

\smallskip

\subsubsection{Analytical results for a drop with the sticking at the ring}

 \begin{figure*}
 \subfigure[]{
\includegraphics[scale=0.8]{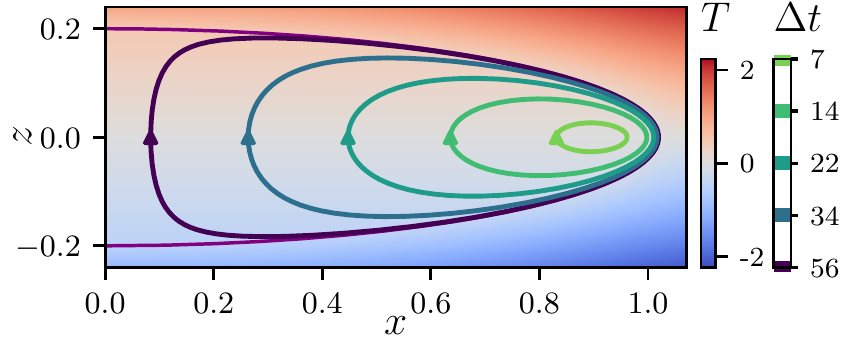} \label{Figure6a}}
\subfigure[]{
\includegraphics[scale=0.8]{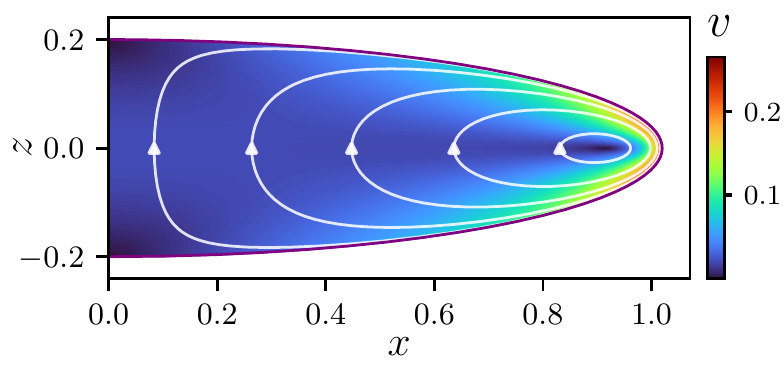} \label{Figure6b}}
  \subfigure[]{
\includegraphics[scale=0.8]{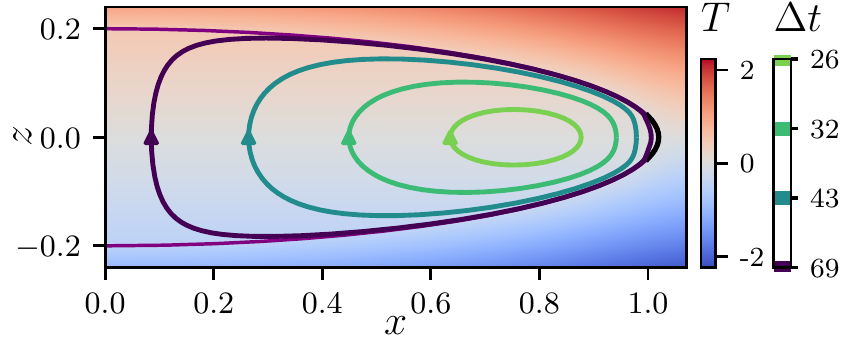} \label{Figure8a}}
 \subfigure[]{
\includegraphics[scale=0.8]{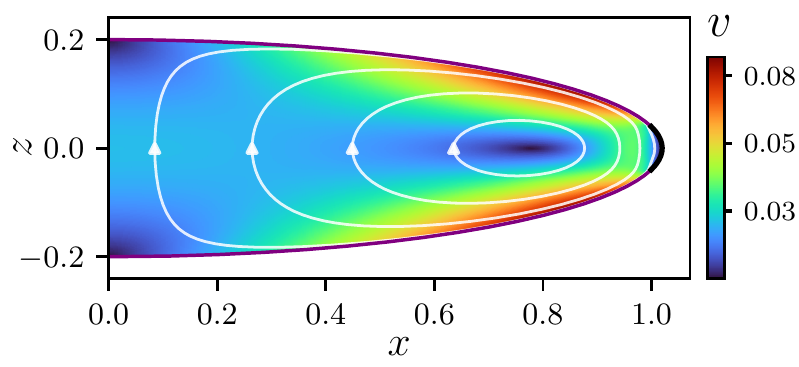} \label{Figure8b}}
\caption{Stream lines,temperature distribution and velocity modulus within  oblate spheroidal drops; \subref{Figure6a} and \subref{Figure6b} -- droplet with a fully free surface ($\xi_0=0.2$,  $\, Ma=1,\, \kappa=0.2$); \subref{Figure8a} and \subref{Figure8b} -droplet suspended on a solid ring ($\xi_0=0.2, \, Ma=1,\, \kappa=0.2$, $u_s = 0.2$). All  the lengths and variables are shown in the dimensionless form. The dimensional values of velocity $v$ and circulation period $\Delta t$  can be deduced  from the dimensionless ones through  multiplication by  the
corresponding scaling parameters: velocity by   $(\chi/H)\hbox{Ma} =  \varsigma A H /\eta$  and period $\Delta t$ by $ (1/\hbox{Ma})\,c/((\chi/H)) = {\eta}/{(2\xi_0\, \varsigma A)}$.}
 \label{Figure8}
\end{figure*}

\begin{figure*}
\subfigure[]{
\includegraphics[scale=0.9]{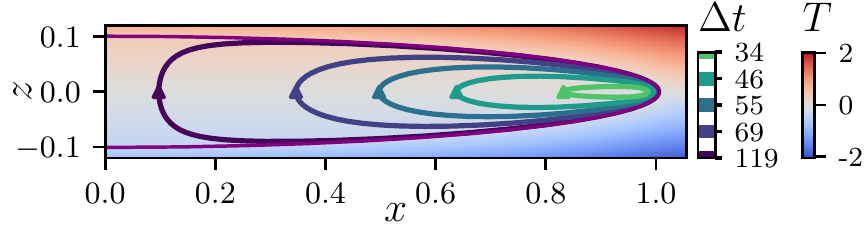} \label{Figure10a}}
 \subfigure[]{
\includegraphics[scale=0.9]{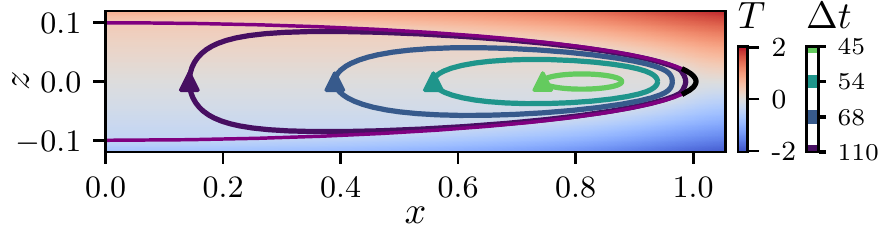}\label{Figure10b}}
 \subfigure[]{
 \includegraphics[scale=0.85]{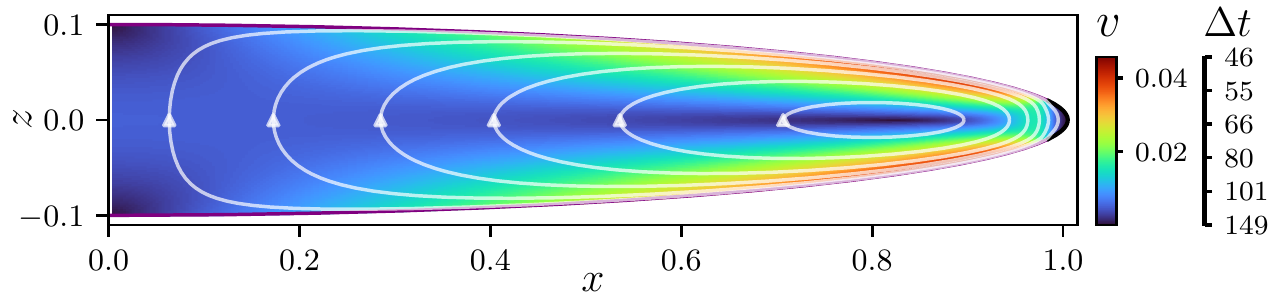}\label{Figure10c}}
\caption{Stream lines, temperature distribution and velocity modulus within oblate spheroidal drops; \subref{Figure10a} droplet with a fully free surface; \subref{Figure10b} and \subref{Figure10c} -- droplet suspended on a solid ring. All the lengths and variables are shown in the dimensionless form
($\xi_0=0.1, \, Ma=1,\, \kappa=0.2$, $u_s = 0.2$).}
\label{Figure10}
\end{figure*}

Let us consider now an oblate drop suspended on a solid ring.
 In this case at the
small part of surface $|u| \le u_s$ near the circular contact line along the drop equator  the  sticking boundary condition is the following:
\begin{equation}
   v_{\tau} = -\, v_u  \propto \,\partial_\xi \psi[u,\xi]\Big\vert_{\xi=\xi_0} = 0 \  \text{ for }  |u|\le u_s \ . \
    \label{vBC}
\end{equation}
The equation means that the tangential velocity component is equal to zero in this region.
  At the same time the boundary condition for the balance of  the tangential viscous and Marangoni forces, Eq. (\ref{xiubc})),  takes place  at the free surface of  an ellipsoidal drop, for which $|u|\ge u_s$.

In order to find the solution of Eq. (\ref{big1}) for the oblate spheroid drop with a suspension ring we apply a set of equations which is analogous to Eqs. (\ref{BcMa}) -- (\ref{ru})  used  earlier in Section C.1.  The difference is that
in addition to the functional, $E^{free}[\{c_j\}]$ (see, Eq. (\ref{BCfr})), we have to introduce a functional, $E^{stick}[\{c_j\}]$, defining the norm of deviation of Eq. (\ref{vBC}) from zero in the sticking area
\begin{equation}
    E^{stick}[\{c_j\}] = \int_{-u_s}^{u_s} \left(\sum_{j>2}^{N_r} c_j \partial_\xi\psi_j (\xi_0, u)\right)^2 \frac{d{u}}{1-u^2} \, . \
     \label{Erst}
\end{equation}
Our aim is to find a set of coefficients $\{c_j\}$ of the expansion of the full stream function over the limited number, $N_r$, of the basic functions $\psi_j$.
This is accomplished by the minimization of the weighted sum of  the deviations provided by the combined functional
\begin{equation}
    E[\{c_j\}] = E^{free} + \Lambda E^{stick}, \; \Lambda>0 \ , \
     \label{Erst1}
\end{equation}
where  $\Lambda$ is a certain undetermined multiplier.
 Likewise the previous case the finite approximation for the stream function with the given accuracy of determination is applied.  The
 inaccuracy  in $E[\{c_j\}]$ defined by  Eq. (\ref{Erst1}) and, therefore, deviations $E^{free}[\{c_j\}]$ and  $E^{stick}[\{c_j\}]$ converge to zero when the number of the basic functions, $N_r$, increases. The parameter $\Lambda$ is chosen to ensure close  contributions from the both mean square deviations (relative deviations of  the norms $E^{free}[\{c_j\}]$ and  $E^{stick}[\{c_j\}]$  from zero are about $10^{-3}$).
As a result an optimized set of coefficients $\{c_j\}_{opt}$ in the expansion for the general stream function, Eq. (\ref{BCsim}),   and  the corresponding final thermocapillary  flow within the drop suspended on the ring  are  obtained, see Figs. \ref{Figure8}\subref{Figure8a}, \ref{Figure8}\subref{Figure8b},  \ref{Figure10}\subref{Figure10b} and \ref{Figure10}\subref{Figure10c}.
 Unlike  the case of a fully free drop, the fastest fluid flow along the vortex trajectory occurs not at the end face points, but at  a certain distance from them. The comparison of Figs. \ref{Figure8}\subref{Figure6b}  and \ref{Figure8}\subref{Figure8b} also  indicates  that the fluid flow is slowing down (about twice) near the clamping area in the drop. This is the way the  sticking (no-slip) boundary conditions along the drop equator affect the circulatory Marangoni flow within the droplets with a suspension ring. However, these differences show up themselves mainly in a butt end region of the drop, and do not  essentially  affect  the convection motion in the main body of the oblate drop. It means  that the shape of the drop's butt end  is not  important for such a case and the approximation of the lens-like shape of  drops by the  oblate spheroid works well.

The above results for a drop suspended on the solid ring are obtained using  analytical calculations based on the governing hydrodynamic equations and the corresponding boundary conditions. However, some important  theory outcomes  can be obtained in a simpler way using certain estimations. For example, one can estimate the characteristic  velocity of the stationary fluid flow that is one of the main characteristics of the convective motion. Let us remind that Marangoni flow is induced by  variations of the surface tension
 $\delta \gamma = -\varsigma \,T$ (see the boundary condition, Eq. (\ref{xiubc})).  The tangential temperature gradient along the free surface of the drop  is finite.
  Thus,  the Marangoni stress is of the order $\varsigma T /R$. On the other hand, the viscous tangential stress in the same area can be written as $\,\eta\, v\,H^{-1}$, where $ v$ and $ H$ are the velocity of the fluid motion and the  corresponding  drop height along the $z$  axes, respectively. From the force balance one obtains the following estimation for the characteristic fluid velocity
  \begin{equation}
 v \sim (\varsigma/\eta) \,(H/R) \, T \, . \
  \label{vchar}
\end{equation}
   One can see that for the fixed values of $T$ (i.e., of the temperature difference across the drop) the characteristic fluid velocity will be smaller for the flattened drops (for which the  ratio $H/(2R)$ is lesser).  This leads to increase of the period of circulation $\Delta t$, see Figs. \ref{Figure5}, \ref{Figure8}, \ref{Figure10}.

The above estimations become  even simpler for the real lens-like drops for which  the curvature radius  is constant everywhere. For such a case the characteristic Marangoni stress on the spherical interface has the same order of magnitude, $\varsigma T /R$, which is compensated by the viscous tangential stress $\,\sim\,\eta v/H$ . By equating these two expressions one obtains the same estimation for the characteristic fluid velocity, $v \sim (\varsigma/\eta) \,(H/R) \, T$, like  for the case of the ellipsoidal drop. Moreover, we can estimate the fluid velocity near the butt end of a drop: ${v_{\mathrm{end}} \sim (H_{\mathrm{end}}/H)\,v}$. Taking into account that ${H_{\mathrm{end}} \ll H}$, one obtains ${v_{\mathrm{end}} \ll v}$. It  confirms our analytical finding that the maximum of the velocity modulus in the oblate droplet on a ring is located at  a certain distance from its butt end, independently of the real shape of a drop (ellipsoidal or lens-like one).

\medskip

\subsubsection{Numerical experiment}

To get further insight about Marangoni flows within oblate fluid
droplets suspended on the circular frame we have  conducted  a  numerical hydrodynamic experiment. The mathematical basis for the numerical simulations, as well as  geometrical and thermal constraints imposed on the drops are analogous to those considered in \cite{Pikina2022}.
A fluid droplet,  suspended on a supporting circular wire, is placed inside a cylindrical chamber between two round plates with different temperatures ($T_\mathrm{up} > T_\mathrm{dn}$) as shown in Fig. \ref{Figure1}. The drop is in the ambient air at the normal atmospheric pressure.
		We assume  that evaporation does not occur, and the droplet shape  does not change over time.
The transfer of heat in the air occurs due to the thermal conductivity. The heat transfer in  a droplet occurs due to the thermal conductivity and convection. We consider only the thermocapillary flow within the droplet, which arises due to the linear dependence of the surface tension of fluid on temperature, see Eq. (\ref{gam}). The material and geometrical  parameters of the drop  are shown in Table \ref{tab:Parameters}.

To simulate a thermocapillary flow within the suspended drops we use the cylindrical coordinates $(r,\phi,z)$. Due to the axial symmetry of a drop the transfer of mass and heat does not depend on the angular coordinate $\phi$. This allows us to consider the hydrodynamic flows in a drop as a two-dimensional problem and proceed with the numerical calculations in the coordinates $(r,z)$ \cite{Pikina2022}. To describe Marangoni flows in a droplet we use  Navier-Stokes
and  continuity equations for  incompressible fluid, and the heat transfer equation,  written in cylindrical coordinates. The above equations are accompanied by a set of  the  appropriate initial and boundary conditions. The hydrodynamic equations are presented in  Stokes stream function-vorticity formulation.
The Marangoni convection problem is solved with the help of  the  commercial package FlexPDE Professional Version 7.18/W64 3D \cite{Liu2018} (for details see \cite{Pikina2022}). The time steps in the program are generated automatically in order to minimize a calculation error. A special attention was also paid to the mesh convergence.

\begin{figure}[!htb]
		\includegraphics[width=0.8\linewidth]{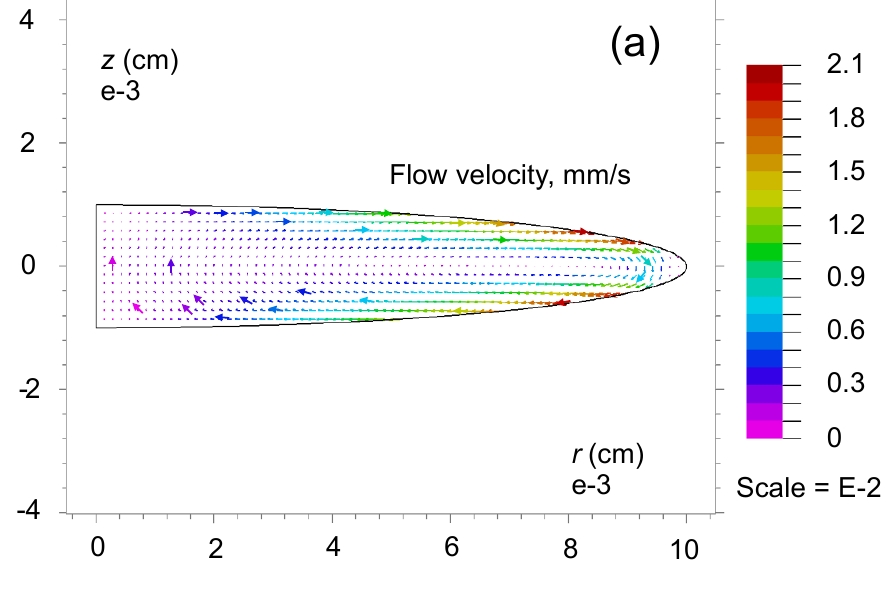} \includegraphics[width=0.8\linewidth]{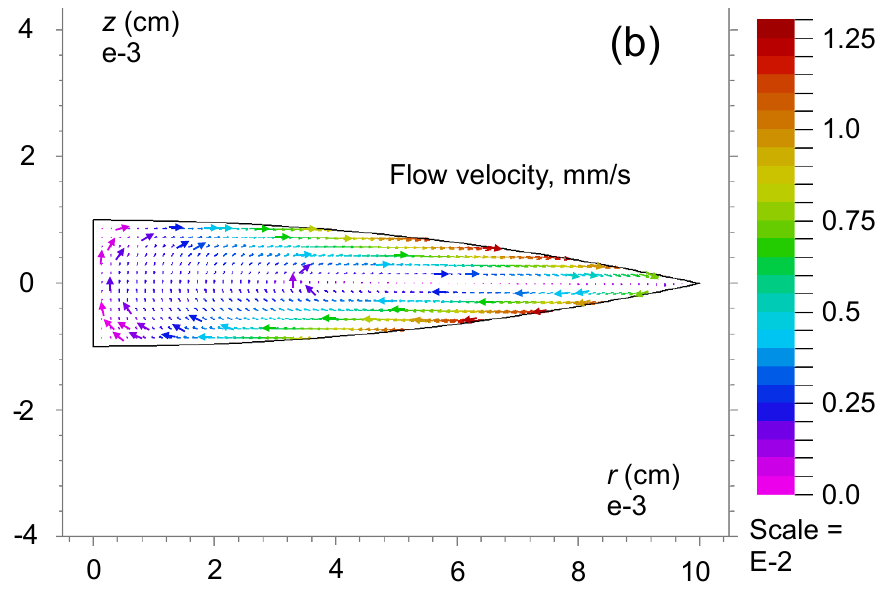}
		\includegraphics[width=0.8\linewidth]{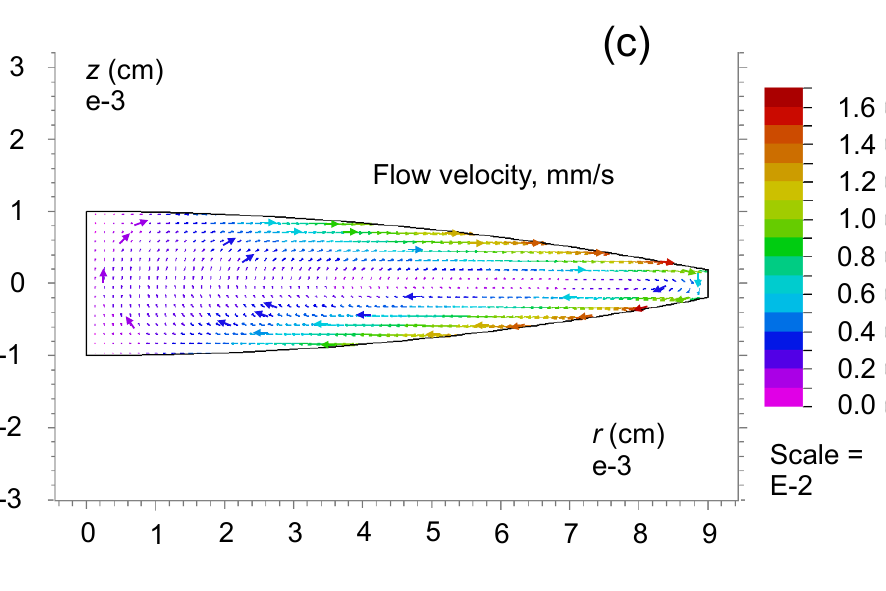}
		\caption{\label{Figure14} The flow velocity field in a drop at  time $t= 20 t_\mathrm{rel}$: (a) ellipsoidal shape, (b) shape of a biconvex lens, and (c) lens with a truncated edge.}
	\end{figure}
		\begin{figure}[!htb]
		\includegraphics[width=0.8\linewidth]{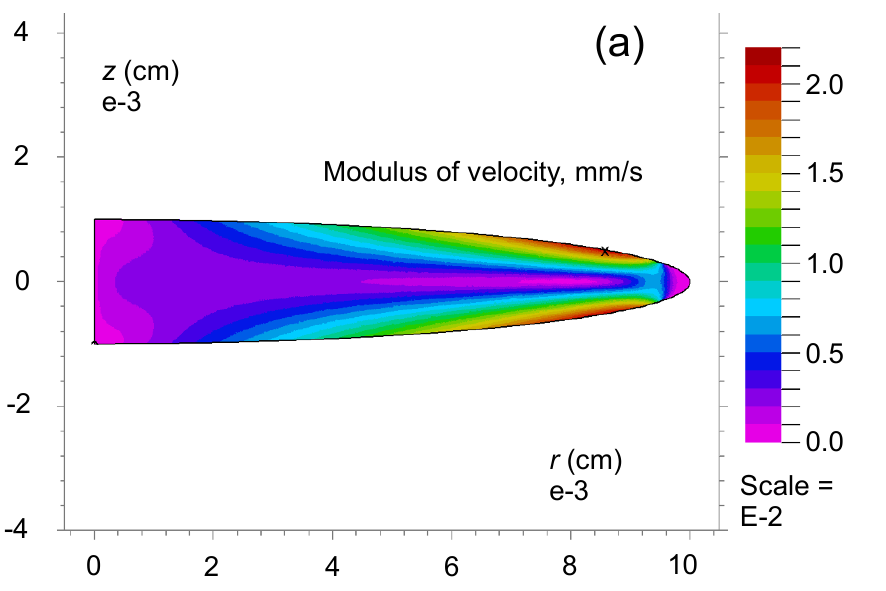} \includegraphics[width=0.8\linewidth]{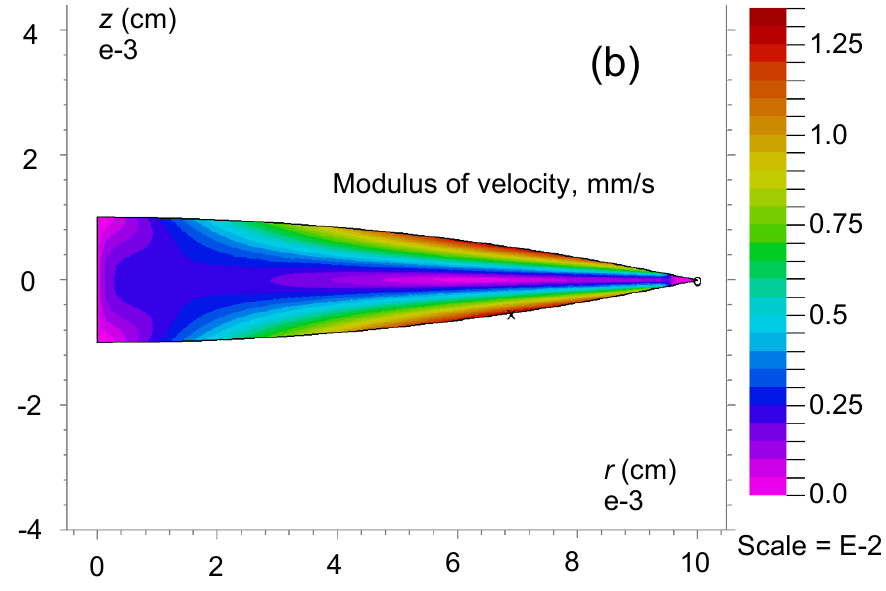}
		\includegraphics[width=0.8\linewidth]{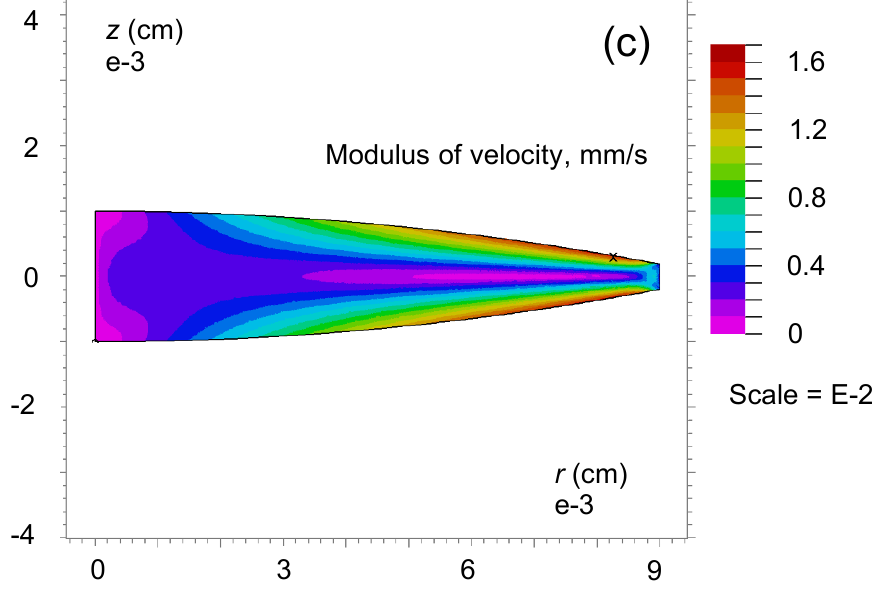}
		\caption{\label{Figure15} The modulus of the flow velocity in a droplet at  time $t= 20 t_\mathrm{rel}$: (a) ellipsoidal shape, (b) shape of a biconvex lens, and (c) lens with a truncated edge.}
	\end{figure}

\begin{figure}
\subfigure[]{
\includegraphics[scale=0.8]{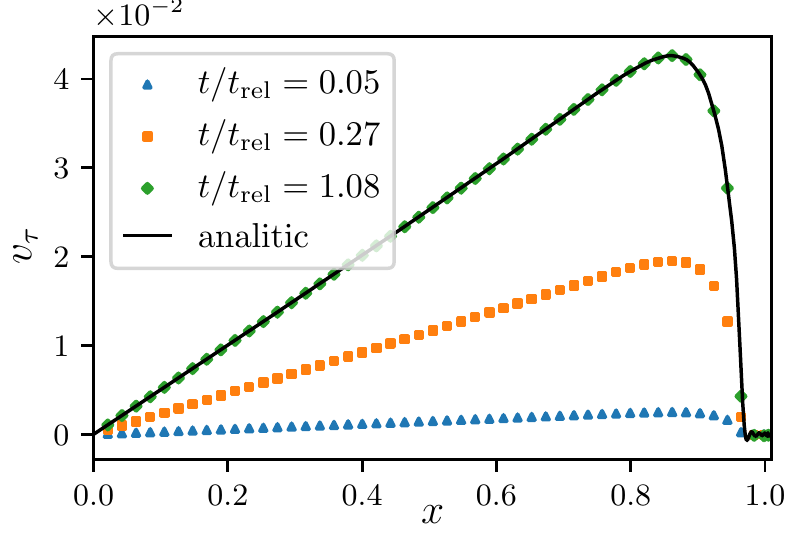}\label{Figure18a}}
\subfigure[]{
\includegraphics[scale=0.8]{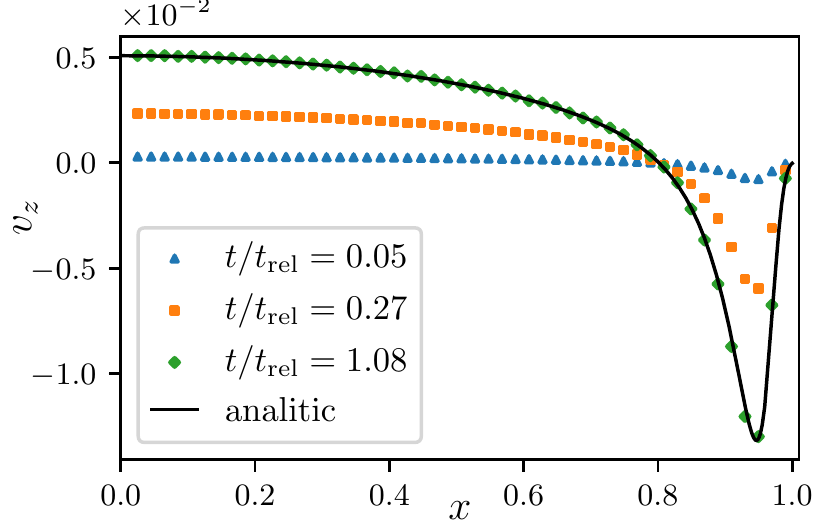}\label{Figure18b}}
\caption{Comparison of  the numerical and analytical results  for the flow velocity components within the ellipsoidal drop on the ring for different relative durations of the numerical experiment; \subref{Figure18a} tangential velocity $v_{\tau}$ (at the upper drop surface)
and \subref{Figure18b} vertical velocity $v_z$ (at $z=0$). Analytical and numerical results are indicated by a solid line and symbols, respectively. For analytical results  $u_s=0.27 ,\xi_0 = 0.1, \mathrm{Ma}=1, N_r = 110$.}
\label{Figure18}
 \end{figure}

\begin{figure}
\includegraphics[scale=0.75]{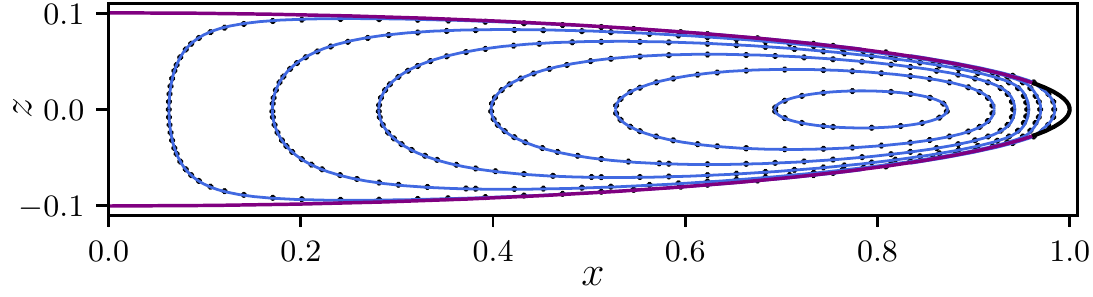}
\caption{Comparison of the numerical and analytical results for the stream function in a case of an
 ellipsoidal drop on the ring. Analytical and numerical results   are indicated by solid line and dots, respectively ($t = 20 t_\mathrm{rel}$).}
 \label{Figure17}
 \end{figure}

 The calculations are performed for the time $t_{max} = 20 t_\mathrm{rel}$ ($t_\mathrm{rel}$ is the heat relaxation time in the air due to the thermal conductivity).
	At the initial  time moment, there is no fluid flow, and the temperature in the entire system is uniform, $T = T_\mathrm{up}$. In a relatively short period of time (less than $t_\mathrm{rel}$  $\approx 0.05$ s), the temperature of the lower plate in the chamber decreases linearly to the value $T_\mathrm{dn}$, and, further
stays constant with time. The values $T_\mathrm{up}  = 334$ K and $T_\mathrm{dn} = 324$ K are chosen for numerical calculations, see Table \ref{tab:Parameters}.
 The temperature value $T_\mathrm{up}$ of the upper plate in the chamber is fixed throughout the entire process.

 The Marangoni boundary conditions, Eq. (\ref{xiubc}) are used for a free surface of the drop. In turn,  the no-slip (sticking) boundary conditions are applied at the border between the liquid and supporting ring ($r > 0.95R$). The corresponding friction  force  turns  the tangential component, $v_{\tau}$, of the fluid velocity to zero. In our numerical calculations  we  consider three types of the drop shape:
  an ellipsoidal drop, a biconvex lens and a lens with a truncated edge. For the later shape, the no-slip (sticking) condition is set at the truncated edge ($r = 0.9 R$).

In Figs. \ref{Figure14}, \ref{Figure15} the distribution of the  fluid flow velocity within a  droplet is shown for different types of the droplet shape. We observe an axially symmetric vortex in  oblate drops of various shapes that  agrees  with our analytical results (Figs. \ref{Figure5}--\ref{Figure10}). The fluid flow circulating in the $(r, z)$ plane is directed clockwise in the case under consideration (Fig. \ref{Figure14}). Because of  $T_\mathrm{up} > T_\mathrm{dn}$,  Marangoni flow is directed along the free surface of the droplet from the hot area to the cold one, (i.e., from the area of the low surface tension to the area of the high surface tension). The fluid flow velocity varies depending on the shape of the droplet (Fig. \ref{Figure14}). Is is due to differences in the curvature of the drop surface and in the shapes of the  contact area with a supporting ring.
The maximum velocity varies from 0.0135 mm/s (for a shape of a biconvex lens) to 0.022 mm/s (for a drop of  an  ellipsoidal shape).

  The  direct comparison of the numerical and analytical results is shown in Figs. \ref{Figure18} and \ref{Figure17}. In Fig. \ref{Figure18} the dependencies of the tangential, $v_{\tau}$, and vertical, $v_z$, components of the fluid velocity on the lateral coordinate $x$ are shown for different durations of the numerical experiment. It is clear  that the system reaches the stationary state pretty fast, within a period that is about the relaxation time $t_\mathrm{rel}$. This ensures that the time chosen for the numerical calculations, $t_{max} = 20 t_\mathrm{rel}$ , is enough to reach the steady state of convection. In Fig. \ref{Figure17} we show a comparison of the stream lines obtained analytically to the ones acquired via the numerical experiment for the ellipsoidal drop on a solid ring. There is a good  agreement between  the numerical and analytical results. Thus, we conclude that the numerical results for Marangoni convection within the oblate droplets on the ring are in good accordance with that obtained by analytical methods.

\section{Marangoni  instability}

In this section we consider the stability of the stationary solutions $(\psi_{st},\, T_{st})$ relative  to the increase of the temperature gradient across the drop.
The analysis is applied to a fully free drop which shows essentially the same features of the critical thermocapillary motion as a drop suspended on the solid circular frame. The differences in the stability criteria, i.e. in the values of the critical Marangoni numbers, Ma$_c$, and in the shape of the critical motion are reduced to  insignificant numerical corrections. It is also true for the role of the drop butt end shape, which  does not influence the circulatory motion in the main body of a drop as it has been detected earlier.
The validity of the  above approach is confirmed by a fact that the critical convection motion is localized close to the central part of a drop and is exponentially small in the end face area (see below). In that follows
 we consider only axially-symmetrical perturbations, which is in accordance with the symmetry of the oblate droplets, and is  technically much easier.

\subsection{ General remarks}

To reveal the stability of the obtained solutions for Marangoni convection  let us consider
  their slight deviations  from those ones of the stationary solutions and analyze  their evolution  in  time.
  While doing this, we introduce general expansions ($\psi_\Sigma = \psi_{st}+\delta\psi, T_\Sigma =  T_{st}  + \delta T$) and substitute them in Eqs. (\ref{big1})--(\ref{big2a}) in order to make a linearization procedure over ($\delta{\psi}, \delta{T}$).
    It is well known that the partial solutions of the  corresponding  linearized dynamic equations can be written  as normal pertubations with exponential  dependence on time \cite{Koschmieder1974,Gershuni1972,Landau6,Lebedev1993,Falkovich,Pikina2022}:
   \begin{eqnarray}
  \delta{\psi} \propto\,\exp[\,\lambda\,t ]      \ , \
  \label{sol1}\\
 \delta{T}\propto\,\,\exp[\,\lambda\,t  ]      \ , \
  \label{sol2}
  \end{eqnarray}
   where parameter $\lambda\,$
       determines the time character of the normal perturbation evolution.
 The stationary solutions for the thermocapillary motion are stable if the condition  $Re[{\lambda}]<0$ is fulfilled for all normal modes; it  corresponds to  the exponential fall off of all perturbations. In general, the perturbations spectrum  depends on  Marangoni number, Ma.
It is quite usual that for  small Ma values all $\lambda_i$  show a negative sign of the real part of $\lambda_i$. However, beginning with  certain larger Ma values  perturbations with the positive $Re[\lambda]$ appear. It  corresponds to the growth of the corresponding perturbations. Thus, the stability loss of a thermocapillary flow is determined by a minimal Marangoni number $\hbox{Ma}_c$ for which
    $Re[{\lambda}]$ reaches the zero value  for the  first time. The corresponding solutions for $\delta{\psi}$ and $\delta{T}$ are named as  critical and   defined below as $(\psi_c, T_c)$.

It is important while doing our calculations we do not overstep  the limit of  the linear Marangoni response approximation, i.e.
 we consider a relatively small temperature gradients across the film and, accordingly, the small Ma values, Ma $\lesssim 10^2$ (see Appendix \ref{sec:A} and Eq.  (\ref{uneqv3}) there).
  As a consequence  we omit the nonlinear terms  $\rho \mathbf{v}_{st}\nabla \mathbf{v}_c$  and $\rho \mathbf{v}_{c}\nabla \mathbf{v}_{st}$ in Navier-Stokes equation, which are related to the momentum transfer by the stationary flux  (see Appendix \ref{sec:C}).
 As a result, the critical velocity  (stream function) field continues to obey the linear equation
(\ref{big1}): $\hat{E}^4 \psi_c = 0$,
and the basis of  the solution for $\psi_c$ remains the same. This allows us to employ the same  expansion $\psi_c = \sum_j c_{cj} \psi_{j}$  over  $\{ \psi_{j} \}$ with a set  of the basic functions,  introduced in Sec. II  A.

\begin{figure}
\includegraphics[scale=0.7]{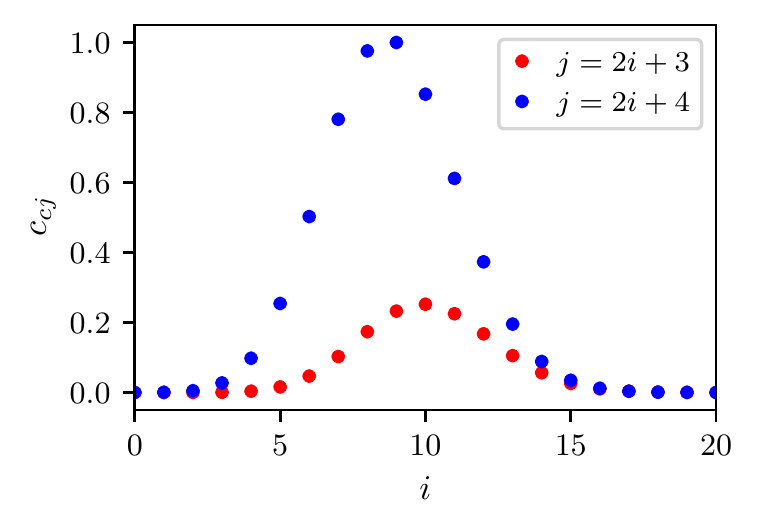}
     \caption{ Coefficients of the expansion of the critical perturbations over  their number $j$  as a function of a current integer number $i$.
     Coefficients for even an odd basic functions  are represented  by blue and red points, respectively.  ($\kappa =0.1, \xi_0 = 0.05$ , Ma$_c = 101.4$).
               }
      \label{cji}
   \end{figure}

\begin{figure*}
\subfigure[]{
\includegraphics[scale=0.85]{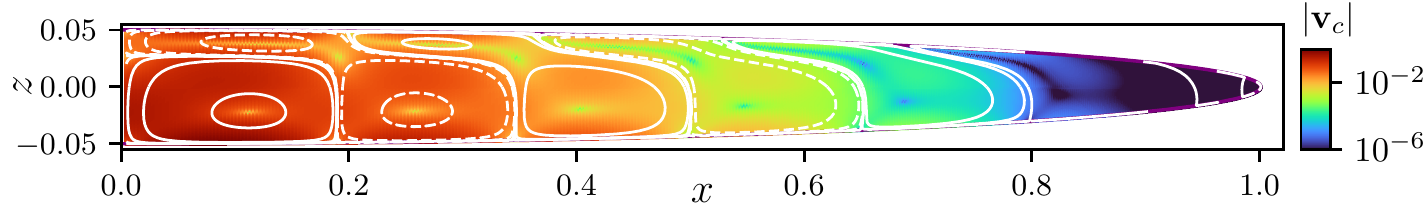}  \label{frCrMa005}}
 \subfigure[]{
\includegraphics[scale=0.85]{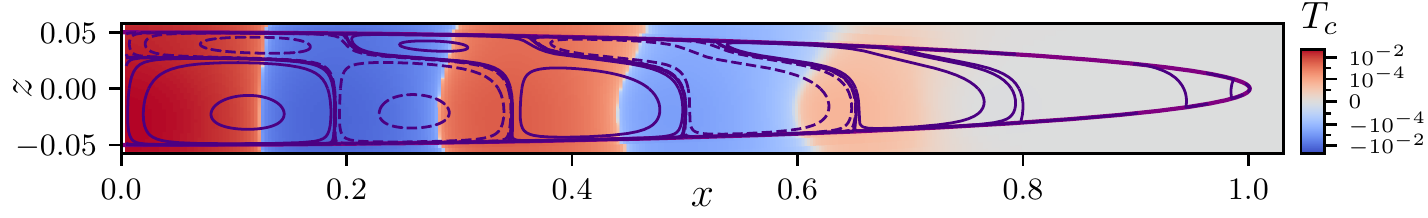} \label{crfree005}}
     \caption{ \subref{frCrMa005}  Critical   convection motion (the velocity modulus is shown in logarithmic scale) within the strongly oblate free drop; \subref{crfree005}  Critical  temperature distribution (shown in logarithmic scale) and convection motion within the free drop:   $\xi_0 = 0.05$, $\kappa=0.1\,$, Ma$_c\approx 101$. The dashed and solid lines indicate the opposite direction of the fluid velocity in the neighboring vortices.}
      \label{crfree05}
   \end{figure*}

\subsection{ Critical Marangoni flow in the axially symmetric oblate drops}

 The heat transport equation $\Delta T_\Sigma = \frac{c}{H}\, (\mathbf{v_\Sigma} \nabla )\, T_\Sigma$ for the critical temperature distribution after linearization takes the form
    \begin{eqnarray}
        \Delta T_c
                = \frac{c}{H}\, (\mathbf{v}_c \nabla)\, T_0
                                 \ , \
         \label{Tc}
    \end{eqnarray}
    where  the term $({c}{H}^{-1})\,(\mathbf{v}_c  \nabla) \,T_c$ is omitted due to a  higher order of smallness.
 We also omit the term in Eq. (\ref{Tc}) which is related to the heat transfer by a stationary flux, i.e.
      the  term  $\frac{c}{H}\,(\mathbf{v}_{st} \nabla)\, T_c$ is neglected in comparison  with $\frac{c}{H}\, (\mathbf{v}_c \nabla)\, T_0$ due to its smallness $\propto\,({H/R})$ in a flattened drop, see Appendix \ref{sec:C}.
      There is also an additional smallness $\propto\,({H/R})^{1/2}$  related
to the fact that the  critical convective motion and the stationary convection within an oblate drop are significantly divided in space (see below). Thereby, the influence of the stationary convective motion on the  critical temperature pertubations distribution  within the drop is not taken into account in further derivations.

        The Eq. (\ref{Tc}) coincides with equation (\ref{Tj}) which allows the temperature amendments of the first order to be calculated as a response to a set $\{\psi_{j}\}$. In such a way  we obtain an expansion $T_c = \sum_j c_{cj} T_{c\,j}$ (compare with \cite{Pikina2022} and see Appendix \ref{sec:B}).

Now, having in hands the analytical expressions for the temperature
distribution within an oblate drop in the critical regime, we can shift to derivation of the critical Marangoni number,
$\hbox{Ma}_c$, and  corresponding vector $\{c_{cj}\}$.
 These variables should  satisfy Marangoni boundary condition (\ref{xiubc})   at  $\xi=\xi_0$:
\begin{eqnarray}
  \Big\{\partial_\xi {\psi}_c \, -\frac{1}{2\xi}(\xi^2+u^2)\,\partial_\xi^2 {\psi}_c\, \Big\}_{\xi=\xi_0}
 \qquad \qquad \nonumber \\
  =   \,- \,\hbox{Ma}_c\,\Big\{\frac{(u^2  +  \xi^2)^{3/2}} {2\,\xi\,\sqrt{1  +  \xi^2 }}\, ({1 -  u^2\,})  \,\partial_{u} T_{c} \Big\}_{\xi=\xi_0}\ , \
\label{BCn}
\end{eqnarray}
where each $\partial_{u}T_{c j}$ is written  in accordance with  expression  (\ref{Tcj}) (see Appendix \ref{sec:B}).
 By analogy with designations introduced in Eqs. (\ref{BCsim}) -- (\ref{ru}) the right part of the equality (\ref{BCn}) can be rewritten
as
 $\hbox{Ma}_c \sum_j \,r_j[u]\, c_{cj}$.
As a result Eq. (\ref{BCn}) reads
\begin{equation}
    \sum_{j>2} c_{cj}\, l_j[u] = \mathrm{Ma}_c \sum_{j>2} c_{cj} r_j[u] \, , \
\label{bcin}
\end{equation}
where
\begin{equation}
r_j(u) =  \frac{-(\xi_0^2+u^2)^{3/2}}{2\xi_0\sqrt{1+\xi_0^2}}\, \,(1-u^2)\partial_{u}T_j(u, \xi_0)\, . \
 \label{rin}
\end{equation}
Likewise to earlier consideration presented in Sec II. C, the boundary conditions, Eq. (\ref{bcin}) are fulfilled when the corresponding mean square deviation turns to zero:
\begin{equation}
E = \int \frac{d{u}}{1-u^2}\Big|\sum_j [l_j(u) - \mathrm{Ma}_c r_j(u)] \, c_{cj}\Big|^2 = \,0\, . \
 \label{Erin}
\end{equation}
Thus, for the fixed  $N_r$ values our minimization procedure is reduced to
  the quadratic eigenvalue problem \cite{QEP} in analogy with our previous work \cite{Pikina2022}.
 Accordingly, the vector $ \{c_{c j}\} $ and  the corresponding  minimal critical value  \hbox{Ma}$_c$
  are determined from the equation (\ref{Erin}) using an algorithm described in \cite{Pikina2022} (see also \cite{ImMa}).

In Fig. \ref{cji}  the  coefficients of the expansion of the solution for $\psi$ over  basic functions  are shown in dependence on their number $j$. This graph illustrates the fact that only a finite number of the  basic functions contribute  to the general critical flow within the oblate drop.

The critical Marangoni motion for the strongly oblate drop (ellipticity ratio $\xi_0 = 0.05$) corresponding to an  expansion presented in Fig. \ref{cji}   is shown in Fig. \ref{crfree05}.
There are several specific features of the critical thermocapillary flow in such drops. First, the number of vortices in the radial drop crossection increases upon the drops flatten out.
Second,  there is an additional  row of the critical vortices (smaller in size) in the upper part of a drop. The reason for their formation is quite clear. According to Fig. \ref{crfree005} there is an alternation of a "hot" and "cold" regions along the lateral direction in the drop.  Let us remind that the boundary Marangoni condition at the drop surface asserts that the fluid flows in the direction from a hot area to  cold one. At the same time in the case of one row of critical vortices there would be a situation when the fluid flows at  one of the drop surfaces  in the opposite direction - from  a cold area to hot one. Such a contradiction is eliminated by  an appearance of a second row of vortices, which restore the correct direction of flow at the whole drop interface.

\begin{figure}
    \centering
    \includegraphics[scale=0.75]{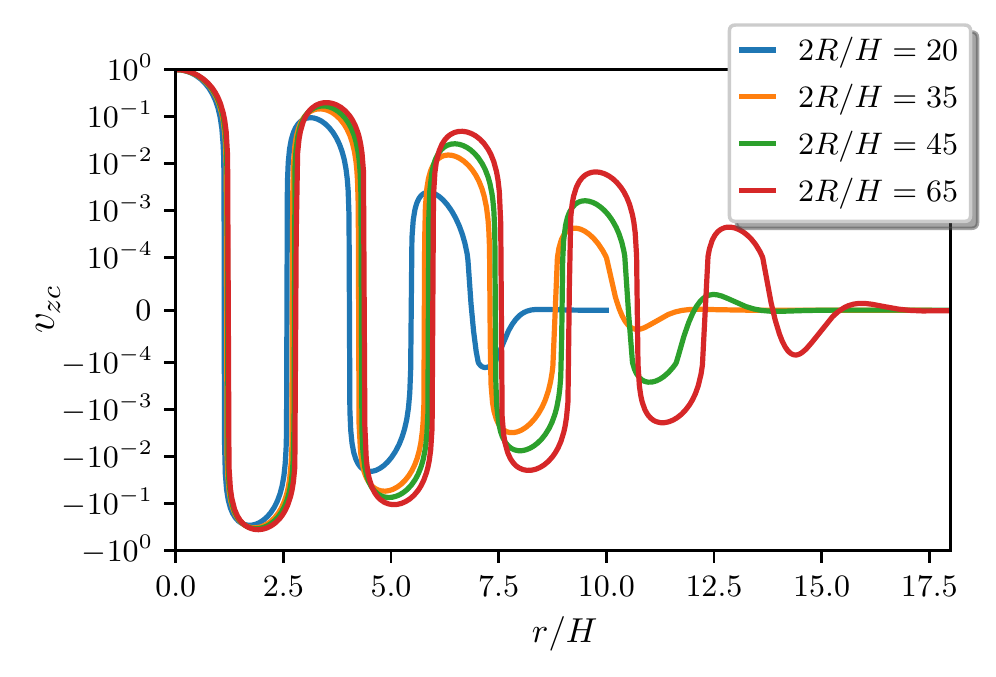}
    \caption{Dependence of  the critical perturbation velocity component $v_{z c}$ in the middle plane of the drop  on the relative radial coordinate $r/H$.
    Note the logarithmic scale for the velocity modulus. The dependence shows an oscillating character with the fast exponential decay of the velocity amplitude. The scale of the exponential fall off  increases with the inverse ellipticity ratio of a drop $\xi_0^{- 1} \approx 2R/H$.
      }
    \label{vc}
\end{figure}

There is one more essential  feature one can observe in Fig. \ref{crfree05}. The critical convection modes are almost not  penetrating  in the butt ends of the drop.
This is a new and exciting result which needs a certain explanation. In  Fig. \ref{vc}, one can see that the velocity  modulus of the critical perturbations has an oscillating character with the exponentially decaying amplitude at a certain distance, $r_c$, from the drop center.
This length determines
 the scale of localization of the critical convection mode within the drop. One can see that the lateral size of localization increases with the radius of the drop for the constant values of $H$.

The localization of the critical convection modes close to the drop's center can be understood as follows.
The current drop height  $h[r] = H\sqrt{1 - r^2/R^2}$ diminishes slowly with increase of a  lateral  distance $r$ from its symmetry axis.
Thus, one can  consider  a drop surface as locally flat and the  critical variables  $T_c$, $\mathbf{v}_c$  as  characterizing the critical flow in a local flat layer of the height  $h[r]$ with corresponding
    Marangoni number
      \begin{equation}
    \mathrm{Ma}_{loc}[r] = \mathrm{Ma}_c \, (h^2/H^2) = \mathrm{Ma}_c \,(1 - r^2/R^2) \ . \
\label{Maloc}
\end{equation}
The above equation is written with account to Eq. (\ref{Ma}) which points on the square dependence of Marangoni number on a local thickness (i.e. $\mathrm{Ma}_{loc}[r] \propto h^2[r]$). According to equation (\ref{Maloc}) $\mathrm{Ma}_{loc}[r]$ diminishes (slowly) upon increase of a radius $r$.
This allows us to determine the characteristic size $r_c$ of the predominant critical Marangoni flow within the drop
from the condition $\mathrm{Ma}_{loc}[r]= \mathrm{Ma}_c^{(fl)}$
 \begin{equation}
     r_c = R \sqrt{1 - \frac{\mathrm{Ma}^{(fl)}_c}{\mathrm{Ma}_c}} \,, \
\label{hc}
\end{equation}
where $\mathrm{Ma}_c^{(fl)}$ is the critical Marangoni number in a flat liquid layer, see Appendix \ref{sec:E}.
 In accordance with Eq. (\ref{hc})
the drop can be divided into two lateral parts:

The first region:   $\mathrm{Ma}_{loc} > \mathrm{Ma}_c^{(fl)}$, i.e. $r<r_c $ -- so-called "allowed"region,
 within  which
 for each drop region with a locally flat surface there are  two  real solutions  of the equation $\mathrm{Ma}^{(fl)}[k_\pm] = \mathrm{Ma}_{loc}$,  where $\mathrm{Ma}^{(fl)}[k]$ is determined by Eq.  (\ref{Ma3}) for a flat liquid layer, $k$ is the  lateral dimensionless wave vector, see Appendices \ref{sec:E}, \ref{sec:F}.

  The second  region:    $\mathrm{Ma}_{loc} < \mathrm{Ma}_c^{(fl)}$, i.e. $r>r_c$ -- so-called
   "forbidden" region, where  the wave vector $k_\pm$ becomes complex. The appearance of the image component in the wave vector  $k$ provides an exponential decay of the critical variables $T_c$, $\mathbf{v}_c$ in the plane of a drop,  see Appendix \ref{sec:F}.

 Such an approach allows us to derive the analytical solutions for the critical velocity and temperature perturbations in the flattened drop of a finite size, see Appendix \ref{sec:F}.
  An important outcome of these calculations is an expression for the critical deviation $\delta \mathrm{Ma}_c$  which is defined by the difference between the critical $\mathrm{Ma}_c$ value for the drop of a finite size and that of the flat liquid layer
  $\delta \mathrm{Ma}_c = \mathrm{Ma}_c - \mathrm{Ma}_c^{(fl)}$,
       see equation (\ref{dMac}):
\begin{equation}
   \delta \mathrm{Ma}_c = \frac{H}{2R} \, \sqrt{\alpha_0 \mathrm{Ma}_c^{(fl)}} \, , \
   \label{mdMac}
\end{equation}
where $\alpha_0 = {\frac{d^2 \mathrm{Ma}^{(fl)}}{dk^2}}\Big\vert_{k=k_c}$, and $k_c$ is a critical dimensionless wave number corresponding to periodic modulation of the convection motion in a flat liquid layer (see  Appendix E). After substitution of the Eq. (\ref{mdMac}) for $\delta \mathrm{Ma}_c$ into Eq. (\ref{hc}) we obtain an  explicit expression  for the
 lateral localization $r_c$ of the critical convection mode (lateral size of the "allowed"region) in a drop of the given geometrical parameters $H$ and $R$:
\begin{equation}
   r_c \,\simeq\, \sqrt{H\,R} \,\left(\frac{\alpha_0 \,}{4\,\mathrm{Ma}_c^{(fl)}}\right)^{1/4} \,\sim \, \sqrt{H\,R} \,. \
   \label{frc}
\end{equation}

\begin{figure}
    \includegraphics[scale=0.6]{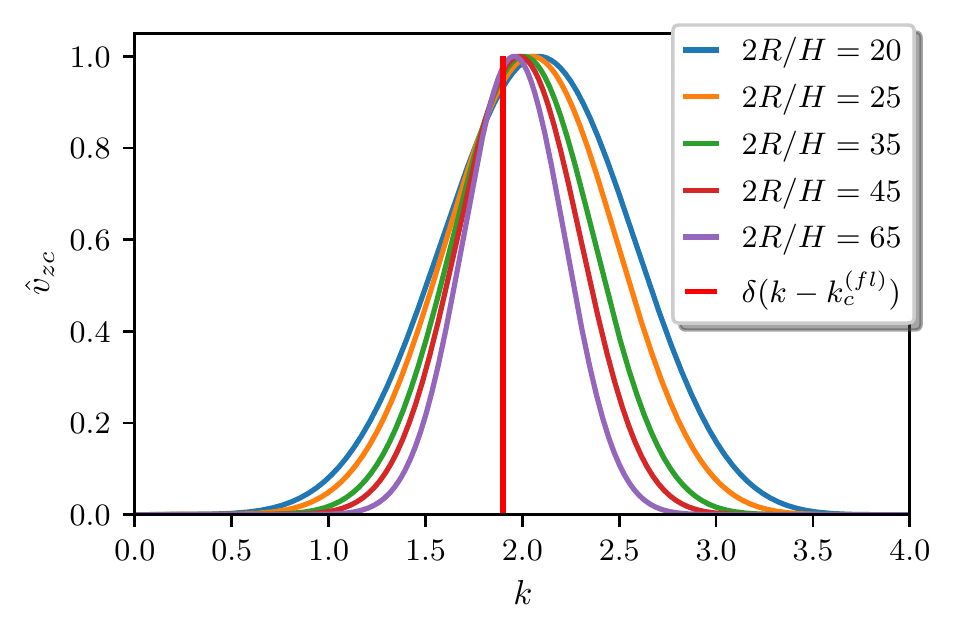}
    \caption{Hankel transformation  of  the critical velocity perturbation  $v_{z c}$ in the symmetry plane $z=0$.
     Separately  the dimensionless critical wave number $k_c = 1.9$ for an infinite flat liquid layer is shown;
      $\kappa=0.1$.}
    \label{fig:hankel_T}
\end{figure}
\begin{figure}
\centering
    \includegraphics[scale=0.7]{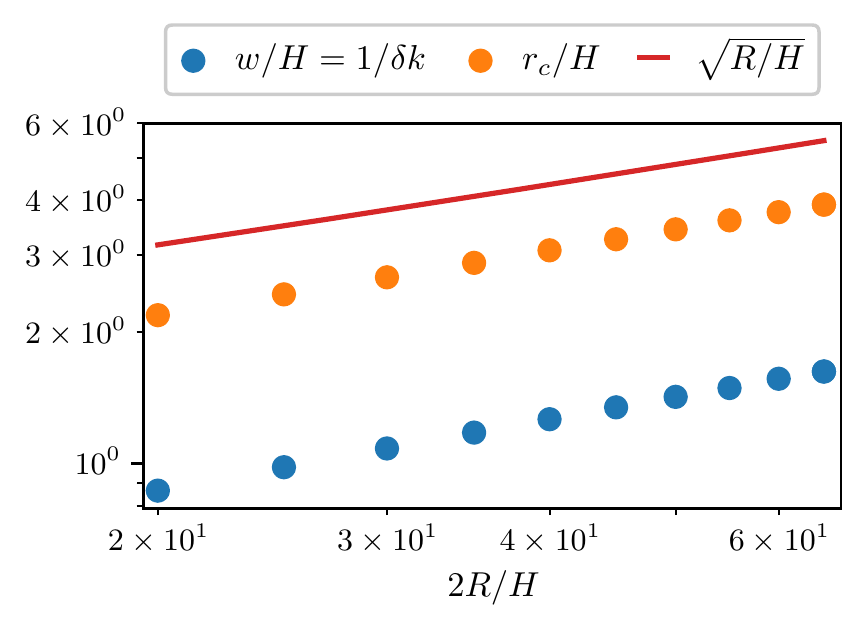}
    \caption{Dependencies of the relative length of localization of the critical mode $w/H=1/\delta k$ (blue dots), and the relative size of the  "allowed" region $r_c/H$ (orange dots) on $2R/H$ ratio. Note the double logarithmic scale. The asymptotic estimation for $w/H$ and $r_c/H$  $\propto \sqrt{R/H}$ is shown by a solid red line.
        }
    \label{fig:w_R}
\end{figure}

\begin{figure}
    \centering
    \includegraphics[width=0.8\linewidth]{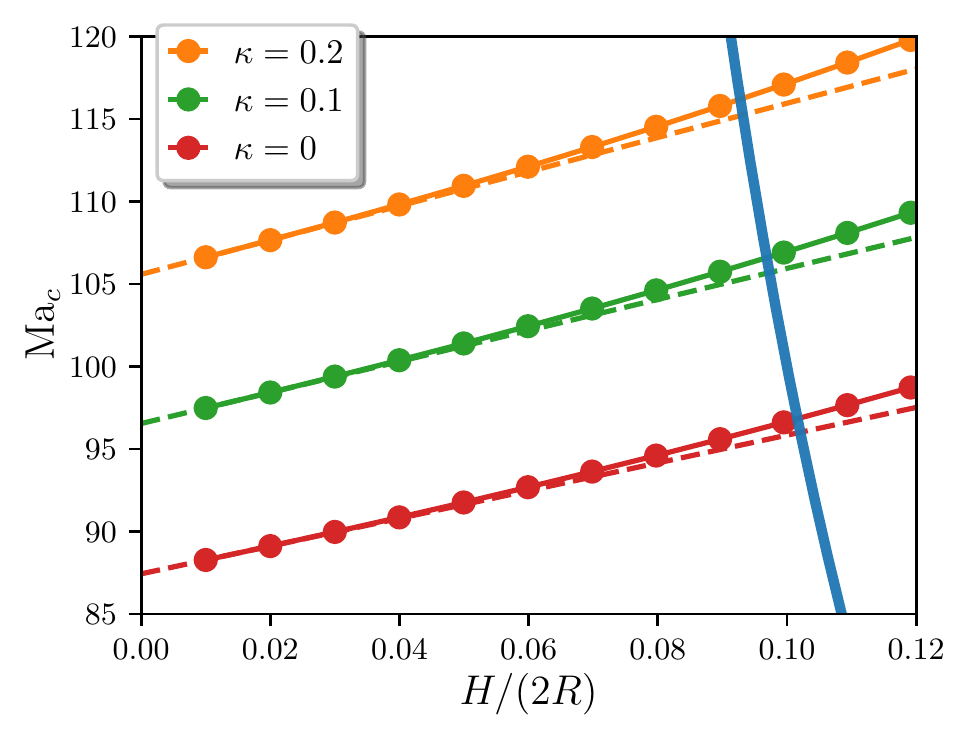}
    \caption{The critical Marangoni number Ma$_c$ as a function of the ratio $H/(2 R)$ for different values of the relative heat conductivity $\kappa $.
        The solid blue line indicates the border between two regimes of the critical Marangoni flow:
          the linear approximation at the far left  below and non-linear one at the upper-right. The dotted lines underline the deviation from the linear Ma$_c$ dependence using Eq. (\ref{mdMac}).
        }
    \label{Mac}
\end{figure}

The above result can be obtained in a somewhat different way. This can be made by analyzing the lateral structure of Marangoni critical convection flow in an oblate drop using Hankel transformation.
 In the case of a flat fluid layer the critical  variables
$v_{cz}, T_c \propto J_0[k_c r/H]$,  where $J_0$ is a Bessel function of a zero order  (see  Appendix E). Hence,  the  critical  Marangoni solutions for an oblate drop can be analyzed using
 the zero order Hankel transformation:
\begin{equation}
    \hat{v}_c[z,k] = \int_0^\infty v_c[z,r] J_0[k r/H] r d{r} \ . \
    \label{Tck}
\end{equation}
The transformation, Eq. (\ref{Tck}) is written for the case of an infinite flat liquid layer. For a drop of a finite radius $R$, the limit of integration should be set equal to $R$. However, this is insignificant for the integral convergence since the exponential decay of the critical Marangoni solution occurs in the region  $r_c\ll R$. For the case of an infinite fluid layer the parameters characterizing the critical motion, $v_c$ and $T_c$, are represented as a delta function at $k = k_c$:
\begin{equation}
    \hat{v}_c^{(fl)}[z,k] \propto \delta[k-k_c] \ , \
    \label{Tcfl}
\end{equation}
 see Fig. \ref{fig:hankel_T}.
For the case of a drop of a finite size the broad peak ("fuzzy delta-function") appears, see Fig. \ref{fig:hankel_T}.
One can see that the peak becomes broader with the diminishing of the lateral drop size $R$. The  peak width in  the $k$-space $\vert\delta k\vert$ (peak half-width at  half-height) is
related to the corresponding length in the real space $w$ as
  $w/H = 1/\vert\delta k\vert$.
The above approach allows us to make an independent estimation of the length of the lateral localization of the critical mode.
 This is done by plotting the
relative lateral length $w/H \sim 1/\vert\delta k\vert$ and the relative size of the allowed region of the critical flow within the drop $r_c/H$ versus $2R/H$,  Fig. \ref{fig:w_R}. Using the double logarithmic scale the validity of the analytic asymptotic result $\,w/H\,\sim r_c/H\sim \sqrt{R/H}$   is directly proved.
 This correlates with the asymptotic result $r_c \propto (HR)^{1/2}$, Eq. (\ref{frc}), obtained for the dimensional localization length, and explains, why the penetration of the critical convection flow in the terminal part of a drop, where $\mathrm{Ma}_{loc} < \mathrm{Ma}_c^{(fl)}$, is troublesome. Thus, in the butt ends of a drop one expects the dominance of a
   stationary convection flow over the critical one. We emphasize that above estimation  of the length of the lateral localization of the critical convection mode $w\sim r_c\sim \sqrt{H R}$ is in accordance with exact solutions of Marangoni convection problem in oblate drop, Fig. \ref{vc}.

The dependence of the critical Marangoni number,  Ma$_c$, on the drop's axes ratio is shown in Fig. \ref{Mac}. The margin between the linear and non-linear regime of the thermocapillary convection is calculated on the basis of the inequalities (\ref{uneqv2}),  (\ref{uneqv3}), see Appendix \ref{sec:A}. One can  see that the linear approximation works fine for the strongly oblate drops (smaller values of the  ratio $H/(2 R)$. This means that
  for  the  sufficiently flatten drops and for the temperature gradients corresponding to the critical Ma  values the transition from the stationary convection to critical one takes place in  a linear regime.
On another hand, the  critical motion for the drops with a large ellipticity ratio:  $\xi_0 \propto H/(2 R)$, occurs  in the strongly nonlinear regime.

\begin{figure*}
    \includegraphics[scale=0.8]{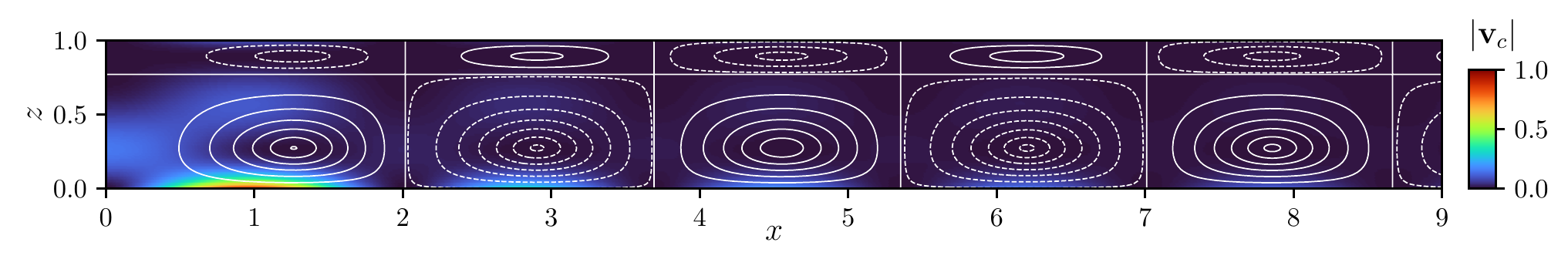}
    \caption{ Critical thermocapillary flow in the limit of a flat film with two free surfaces  ($\mathrm{Ma}_c\approx 87.5, k_c\approx 1.9$). The dashed and solid lines indicate the opposite direction of the fluid velocity in the neighboring vortices.
}
    \label{freelay}
\end{figure*}

 \subsection{Crossover to a flat fluid layer and possible experimental observations}

The above approach to description of Marangoni convection in the  oblate droplets provides a unique possibility to study a crossover from the ellipsoidal drop to a flat liquid layer with two free interfaces. This is made by reducing of the droplet ellipticity ratio $\xi_0\approx H/(2 R)$ to zero, thus leading in the limit $\xi_0\to 0$  to the case of a flat film.
The values of the critical Marangoni number Ma$_c$ and corresponding critical dimensionless wave vector $k_c$ in the limit ($\xi_0\to0, \kappa\to0$ ) constitute  Ma$_c \approx 87.5$  and $k_c \approx  1.9$. The axially symmetric critical convection flow for a flat liquid layer is shown in Fig. \ref{freelay}. The critical wavenumber $k_c$ determines the inverse size of the arising convective cells in the horizontal plane of the film. It should be noted that the correct  value of the critical Marangoni number for a flat fluid layer is about two times larger than that indicated in our earlier paper \cite{Pikina2021}. This is due to a misprint in above paper; more detals are presented in Appendix E.

As we show in Appendix \ref{sec:E} the solution for the critical Marangoni motion is proportional to Bessel function of the zero order $J_0[r]$. This function is characterized by the power law decay in the limit of the large arguments  ($J_0[r] \propto 1/\sqrt{r\,} $ at $r\to \infty\,$, \cite{NLebedev65}). Because the characteristic scale of the critical convective mode increases with the base radius of a drop, this suggests that in the limit of a flat liquid layer the exponential decay of the magnitude of the critical variables $(v_c, T_c)$ is replaced by the power law one. Thus, in the limit of a flat liquid layer the fall off of the variables characterizing the critical Marangoni flow has a gradual power law asymptotic behavior,  Fig. \ref{freelay}.

As follows from the obtained critical convection patterns and instability curves an increase in the drop sphericity leads to a growth of the critical Marangoni number. As we have shown earlier \cite{Pikina2022} the same effect takes place also for a drop with the sticking (no-slip) conditions at the  bottom interface. It is interesting to mention that although the reason for the thermocapillary instability is the presence of the free surface in a drop, the threshold for  critical Marangoni flow for a  fully free drop is higher than for a drop with the sticking conditions at the bottom interface.

	The above theoretical findings can be checked in thermocapillary experiments on the oblate droplets of various size and ellipticity ratio. The most important quantitative characteristic of  Marangoni vortices in the drops that can be directly measured in experiment is time period, $\Delta t$, i.e. the time interval required for the movement along the closed vortex line. The calculated $\Delta t$ values for variant drops are presented in Figs. \ref{Figure5}--\ref{Figure10}. As regards the critical Marangoni flow,  the most intriguing experiment is an  observation of the lateral separation of the critical and stationary convection motion within the drops. In order to visualize this the small overcritical regime should be used. That is the temperature gradient accros the drop should correspond to Ma values exceeding Ma$_c$ for a given drop geometry by about 10$\%$.
In this situation we expect to observe the localization of the many critical vortices near the central part of a drop (in the region $r < r_c \propto \sqrt{HR}$). In the remaining part of a drop, closer to its butt end, the existence of a single stationary vortex is expected.
The first experiments in this direction were carried out in recent thermocapillary study of the flattened sessile drops on the substrate \cite{Zhu2019,Shi2017}.

\section{ SUMMARY}

We have developed a hydrodynamic theory of Marangoni convection in the axially symmetric oblate liquid droplets using the formalism of the Stokes stream functions. Due to the nonuniform temperature distribution the tangential Marangoni force induces a fluid flow along its curved interface, making the thermocapillary flow within the drop thresholdless. Both a drop with the fully free surface and a drop suspended on a solid ring with the sticking boundary conditions have been analyzed. In both cases the analytical solutions predict a stationary thermocapillary flow within the drop in a form of torroidal-like vortices.
However, there is a principal difference in  the thermocapillary flow for the drops with a different type of confinement. For a fully free drop the fastest fluid flow along the vortex trajectory occurs at its end face, while for a drop on a ring the fluid velocity maximum is significantly shifted from the drop's butt end. Moreover, the calculations indicate  subtle slowing down of the fluid velocity close to the bounding ring. Nevertheless, the shape and dynamics of vortices in the drop interior are not seriously affected.

 The general stream function and velocity fields are derived in the stationary regime for the fixed temperature gradients  in dependence on the droplet ellipticity ratio and heat conductivity of the fluid and air. In parallel, numerical experiments on the thermocapillary flow in oblate droplets suspended on the solid frame are carried out. Three types of drop shapes including the oblate spheroid, biconvex spherical lens and lens with truncated end have been modeled. The results of the numerical calculations for the drops of the ellipsoidal shape are in good agreement with that obtained by means of  analytical derivations.
  Moreover,  the comparison of the   solutions for ellipsoidal drops and ones in a shape of biconvex spherical lens indicates  that under sticking boundary conditions at  a drop equator  the convection flow for  these geometries  coincides in the drop interior. This justifies the validity of approximation of a lens-like drop by an  oblate spheroid.

 We have calculated the limitations on the magnitude of the temperature gradient across the drop, which determine the applicability of the linear approximation of the perturbation theory. Within these limitations the corrections to the temperature distribution in the drop due to the  convection motion can be disregarded.

	The solutions  for the temperature distributions and velocity fields corresponding to  the critical regime of Marangoni convection are obtained for the large temperature gradients across the drop.
Perhaps, the most important result obtained here is the lateral separation of the critical and stationary solutions within the drops. The critical vortices are localized  near the central part of a drop (in the region $r < r_c  \propto \sqrt{HR})$,
while the intensive stationary flow is located closer to its butt end.
The role of the critical convection motions is getting especially noticeable under condition that the convection corrections to the temperature distribution within the drop hold (the case of the  large Ma numbers).
	Finally, a crossover to the limit of a flat fluid film is studied. Such a transformation is made by reducing the droplet ellipticity ratio to zero value. It is shown that for such flattened droplets and under an action of the considerable temperature gradients  the formation of a series of critical vortices distributed within the plane of a drop takes place.


\section{No conflicts of interest}

The authors have no conflicts to disclose.

\section{Acknowledgments}
We are grateful to Vladimir V. Lebedev, Efim I. Kats,  Igor  V. Kolokolov and  Sergey S. Vergeles for  fruitful discussions.
The contribution of the  scientists  E.S.P. and M.A.Sh. of the Laboratory  "Modern  Hydrodynamics", created in frames of Grant No. 075-15-2019-1893 of  the Ministry of Science and Higher Education of the Russian Federation  in Landau Institute for Theoretical Physics of the RAS, connected with the  general theory of Marangoni convection in isotropic drops suspended on the ring and corresponding   calculations was supported by the Russian Science Foundation (Grant No. 23-72-30006).
Numerical experiment (by K.S.K.)  was supported by the Russian Science Foundation (Grant No. 22-79-10216).
The work on the derivation of the  stress tensor and expressions for the tangential forces in ellipsoidal coordinates  and  elaboration of  the thermocapillary experiments (S.A.P. and B.I.O.) was supported by the Ministry of Science and Higher Education within the corresponding State assignments of  FSRC "Crystallography and Photonics" RAS.
 The contribution of the scientists  of the Laboratory  "Modern  Hydrodynamics" E.S.P. and M.A.Sh., connected with the  problem  statement and  solving  the problem of the streamfunctions within the ellipsoidal isotropic drops,  was supported by the Ministry of Science and Higher Education within the corresponding State assignment No. 0029-2021-0003 of Landau Institute for Theoretical Physics of the RAS.

\medskip

\section*{Authorship contribution statement}
E.S.Pikina: conceived of the presented idea, calculated  the Marangoni convection, solved of the problem of the temperature distribution, discussed the results, the final manuscript writing.
 M.A. Shishkin: calculated  the Marangoni convection, developed the original operator method for calculation of the stream functions, discussed the results.
 K.S. Kolegov: made the numerical experiment, discussed the  results.
 B.I. Ostrovskii: conceived of the presented idea,  presentation of the results of the calculations, the elaboration of the thermocapillary experiments,  discussed the  results, the final manuscript writing.
 S.A. Pikin: conceived of the presented idea, worked on the derivation of the  stress tensor and expressions for the tangential forces in ellipsoidal coordinates, contributed to the calculations, discussed the  results.

 All authors read and agreed on the final text of the paper.

  \numberwithin{equation} {section} { \bf
\appendixname{}}

\appendix

\section{Elliposoidal coordinates. Limitations of a linear Marangoni response approximation}
\label{sec:A}

 \begin{figure}
    \hskip-0.1true cm
   \includegraphics[width=0.8\linewidth]{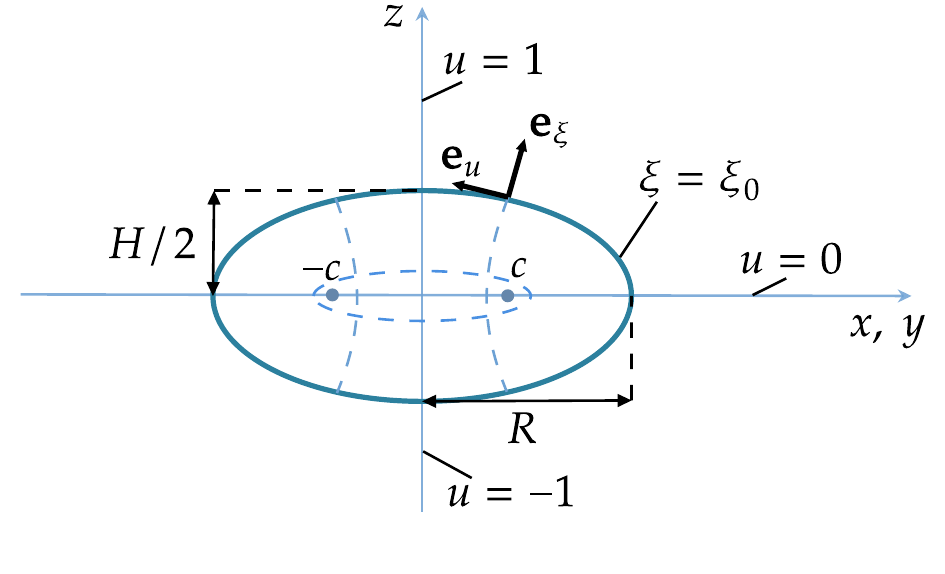}
    \caption{Sketch of the frontal cross-section of a drop in a shape of an oblate spheroid. For convenience, the zero of the $z$-coordinate axis is fixed in the center of the drop.  The   semiaxes of  ellipsoid are $c\sqrt{1+\xi_0^2}\equiv R$ and $c\,\xi_0\equiv H/2$, $\xi_0/\sqrt{1+\xi_0^2}=H/(2R) \ll\,1$ (this inequality is valid for $\xi_0 \ll 1$), $c$  is a focus distance; ${\bf e}_{\xi}$ ,   ${\bf e}_{u}$  are the unit vectors in the oblate spheroidal coordinates in their meridional plane; ${\bf e}_{\xi}$  is outward normal vector to the oblate spheroidal surface of constant ${\xi}=\xi_0$,  unit vector ${\bf e}_{\varphi}$ is the  azimuthal unit vector, oriented beyond the page (sheet) plane, ${\bf e}_{u}$ lies in the tangent plane to the oblate spheroid surface and completes the right-handed basis set $\{{\bf e}_{u}, {\bf e}_{\xi}, {\bf e}_{\varphi}\}$.}
      \label{FigureAp1}
    \end{figure}

We start here with clarifying of the geometrical sense of the oblate spheroid orthogonal coordinates introduced  in Eqs. (\ref{1elc})--(\ref{aLamexi}), see Fig. \ref{FigureAp1}. Every point of space is described by a triple of numbers ($u, \xi,  \varphi\,$), corresponding to a unique point in the Cartesian coordinates $(x, y, z)$. The corresponding orthogonal system of surfaces consists of oblate spheroids formed by surfaces of constant $\xi$ ($\xi = \xi_0$ is a spheroid of the given boundary), one-sheeted hyperboloids of revolution of constant $\vert u\vert$, and planes of $\varphi = const$, where $\varphi$ is an azimuthal angle (for details see \cite{Pikina2022} and references therein).

Next, we consider the limitations of  a linear system of equations describing the  thermocapillary  convection in an oblate liquid drop in linear approximation.

 To justify   the conventional linear  over $T$ thermal energy transport equation in moving fluid
 the unequality
   $\vert \alpha\,\delta p \vert \ll \vert \beta\,T \vert \ll 1 $ should be fulfilled.
   All designations are taken from the density expansion $ \rho = \, \rho_0\,\big( 1 - \beta\,T + \alpha\,\delta p\big)$,
   where $\beta$ and $\alpha$ are coefficients of  the thermal expansion and isothermal compressibility,  respectively,  and $\delta p$ is a  pressure deviation  \cite{Koschmieder1974,Gershuni1972,Pikina2021}. The above inequality is valid due to a fact that the
   fluid velocity is much smaller than the sound velocity ($ \, \simeq 1400\, \text{m/s}$).
We consider  here the temperature range $T \ll 10^3 \text{K}$,
for which the density expansion $\rho = \rho_0(1 - \beta T)$ is certainly valid.

The modified Navier-Stokes equation  and the continuity equation for the incompressible fluid in designations of Sec. II have a form   \cite{Koschmieder1974,Gershuni1972,Pikina2021}:
\begin{eqnarray}
    \frac{\partial {\bf v}}{\partial t} + ({\bf v}\,\overrightarrow{\nabla})\,{\bf v}
   =   - \frac{1}{\rho_0}\,({\bf \nabla} {p})
     + \nu\,\nabla^2 {\bf v} -  \beta\,T\,g\,{\bf e}_z
      \  , \  \  \  \label{NSt01}
      \\
      (\mathbf{\nabla} \,{\bf v})  \,=\, 0
      \ , \qquad \qquad  \qquad \qquad \qquad \qquad \label{cont2}
 \end{eqnarray}
   The term $-\,\beta\,T\,g\,{\bf e}_z$   in Eq. (\ref{NSt01}) corresponds to the convective buoyancy  force in a drop (g is a gravitational acceleration).
 Keeping in mind that Marangoni convection prevails over the buoyant convection
  for drops with  small heights \cite{Koschmieder1974,Gershuni1972,Pikina2021,Pikina2022}
\begin{equation}
H \ll H_c = \sqrt{\frac{\varsigma}{\rho\,g\beta\,}} \sim 10^4 \mu m \  , \
\label{RMa1}
      \end{equation}
  we can  neglect buoyancy   term in Navier-Stokes equation (\ref{NSt01}) in comparison with the viscous term \cite{Gershuni1972,Pikina2022}.
For the ordinary fluids  the transition to buoyancy-dominated convection occurs around a 1 cm, which is many orders of magnitude larger than the droplet heights considered in our theory.

In accordance with calculations made in Sec. II and using continuity condition $\nabla \mathbf{v} =0$, the following estimations for the velocity modulus, $v$, and velocity component, $v_z$, can be obtained:
    \begin{eqnarray}
  v \sim \frac{\varsigma}{\eta} \frac{H}{R}\, T \,, \; v_z \sim v  \frac{H}{R} \ . \
  \label{uneqv}
                  \end{eqnarray}
   One can check that the continuity equation  (\ref{cont2})  is  fulfilled in this case. From the full continuity equation $\nabla(\rho \mathbf{v}) =0$  it follows that
$ (\mathbf{v}\nabla) \delta \rho \,\ll \rho_0 \partial_z v_z$ and
  $\delta v_z \sim v  (H/R)\, \beta T\, \ll v_z$.

Moreover, for the fluid flow with a small velocity, which is characterized by
 Reynolds number Re $\ll 1$
the nonlinear term  $(\mathbf{v}\nabla) \mathbf{v}$ in Eq.(\ref{NSt01}) is negligibly small and can be disregarded in comparison with the viscous term. This  justifies the using of the linearized Navier-Stokes equation.
The smallness of the Reynolds number (i.e.  $\rho_0 (\mathbf{v}\nabla)\mathbf{v} \ll \eta \Delta \mathbf{v}$),
 indicates that the velocity modulus
    \begin{eqnarray}
     v \,\ll  \frac{\eta}{\rho_0\,H} \frac{R}{H} \ . \
     \label{uneqv1}
                  \end{eqnarray}
        Using Eq. (\ref{uneqv})    we obtain from    the inequality (\ref{uneqv1})
         \begin{eqnarray}
        T \ll \frac{\eta^2 }{\rho_0 H \varsigma } \Big(\frac{R}{H}\Big)^2 \, . \
     \label{uneqv2}
                  \end{eqnarray}
                                     Using the definition of Ma, Eq. (\ref{Ma}), one obtains
  \begin{eqnarray}
       \mathrm{Ma}\ll \frac{\eta}{\rho_0 \chi} \Big(\frac{R}{H}\Big)^2 \, . \
 \label{uneqv3}
                  \end{eqnarray}
Taking into account that for the ordinary liquids ${\eta}/{(\rho_0 \chi)} \sim  10^3$, and the ratio $R/H$ for the oblate drops is about $10^1$, we obtain Ma $ \ll 10^5$, which is fulfilled for a Ma range considered in our paper.

 The conventional  thermal energy transport equation in a moving fluid  in a linear approximation over $T$
 can be written as \cite{Gershuni1972,Landau6}):
\begin{eqnarray}
      \frac{\partial {T}}{\partial t}\,+\,({\bf v}\,\overrightarrow{\nabla})\,{T}  \,=\,\chi\,\Delta T  +\, \frac{1}{\rho_0\,c_p} \,\sigma'_{ik} \partial_k v_i  \ , \  \
           \label{apt}
\end{eqnarray}
where $\chi\,=\varkappa\,(\rho_0\,c_p)^{-1}$ is a coefficient of the  temperature conductivity and $\sigma'_{ik}$ is a viscous tensor.
           The term proportional to the fluid velocity in  Eq. (\ref{apt}) can be omitted if the inequality
 $ \vert(\mathbf{v}\nabla)\vert T\,\ll\,\chi \vert \Delta T \vert$ is fulfilled, i.e.
under condition
\begin{eqnarray}
   T\ll \frac{\chi \eta}{\varsigma H} \Big(\frac{R}{H}\Big)^2  \ . \
   \label{uneqvT4}
\end{eqnarray}
In terms of Marangoni number, Ma, this is equivalent to
   \begin{eqnarray}
      \  \mathrm{Ma} \ll (R/H)^2\ . \qquad \qquad
 \label{uneqv4}
\end{eqnarray}

In turn, the dissipation term $\sigma'_{ik} \partial_k v_i$  in Eq. (\ref{apt}) can be disregarded under
condition
\begin{eqnarray}
   \eta v^2  \, \ll \,    \varkappa T \ , \ \qquad \qquad \label{uneqv5}\\
    \implies T \ll \frac{\varkappa \eta }{\varsigma ^2}\, \Big(\frac{R}{H}\Big)^2  \ . \
    \label{uneqvt}
\end{eqnarray}
The above inequalities justify the validity of the  Eqs. (\ref{T01}), (\ref{T0air}).

Thus, we conclude that the inequalities (\ref{RMa1}), (\ref{uneqv2}), (\ref{uneqv3}), (\ref{uneqvT4}), (\ref{uneqv4}) and (\ref{uneqvt}) are fulfilled for the material and transport parameters of the ordinary liquids and for the chosen geometrical characteristics of the oblate drops. These inequalities are also in accordance with the realistic values of the external parameter of a system - temperature gradient across the drop (Ma numbers) used in the  paper. All these give us a confidence that the  linear Marangoni response approximation can be safely applied to the solution of the thermocapillary convection problem in the  oblate droplets. It is needless to say that the above limitations are well within the conventional Boussinesq approximation \cite{Koschmieder1974,Gershuni1972,Pikina2021,Pikina2022}.

\section{Some results on the temperature distribution within the drops}
 \label{sec:B}

 In accordance with  Appendix \ref{sec:A} we can disregard the convectional heat exchange  and the dissipation term in thermal energy transport equations within the drop and in surrounding air, see inequalities (\ref{uneqvT4})--(\ref{uneqvt}).
   As a result, the temperature distributions $T_0, T_{0\, air}$, can be derived from Eqs. (\ref{T01}), (\ref{T0air}) with account to the boundary conditions Eqs. (\ref{bcT0})--(\ref{bcT2f}). The corresponding solutions were obtained in \cite{Pikina2022} and can be written as
    \begin{eqnarray}
 T_{0 \,air} \, = \,  C_{air} \,c
      \underbrace{u \,\xi}_{P_1 \, \Xi_1} +  \alpha_1\, P_1[u]\,\Xi_{1}^{(a)}[\xi] \ , \
    \label{solT1}
\end{eqnarray}
\begin{eqnarray}
T_{0} =  \beta_1 P_1[u] \Xi_{1}[\xi] \ , \
\label{solTd0}
\end{eqnarray}
where
\begin{eqnarray}
 \beta_1 = c \,\underbrace{C_{air} \,\kappa\, \frac{1-\xi_0 (\ln{\Xi_1^{(a)}})'\vert_{\xi=\xi_0}}{1- \kappa \xi_0 (\ln{\Xi_1^{(a)}})'\vert_{\xi=\xi_0}}}_{A} = c\,A,
  \label{BCTtb} \\
   \alpha_1 = \,-\, C_{air}\,  \frac{(1 - \kappa )}{( \Xi_1^{(a)}/\xi_0 - \kappa\,\partial_\xi\, \Xi_1^{(a)})\vert_{\xi=\xi_0} }   \ , \ \
 \label{BCTt}
\end{eqnarray}
 and $\kappa = \varkappa_{air}/\varkappa$ is a relative heat conductivity. The
  functions $\Xi_n^{(a)}$ were derived in \cite{Pikina2022} as the  basic  temperature  distributions  in the air
 and can be written as the linear combinations over
$\Xi_n$, damped at $\xi\to+\infty$:
\begin{eqnarray}
\Xi_0^{(a)}  = - \arctan[\xi] + \frac{\pi}{2} \ , \
    \nonumber \\
\Xi_1^{(a)}  =  \xi \arctan[\xi] - \frac{\pi \xi}{2} -  1 \ , \
   \label{lapl2b}
    \end{eqnarray}
and etc.

   As regards the critical solutions, we note that  Eq. (\ref{Tc}) coincides with equation  $\chi  \Delta  T_1 = v_z \partial_z T_0\,$ for the small temperature deviation  $T_{1}$  from the   temperature distribution $T_0$ (for details see \cite{Pikina2022}).
  This fact allows us to find the temperature response to a set of stream functions $\{\psi_{j}\}$:
  \begin{equation}
    \Delta T_{cj} =  \frac{1}{4\xi_0^2} v_{j}^{z} \, , \
    \label{Tj}
\end{equation}
which can be rewritten in the oblate spheroid coordinates
\begin{equation}
[\hat{\Xi}+\hat{P}]\, T_{cj} = \frac{1}{4\xi_0^2} [u\partial_u - \xi \partial_\xi]\psi_{j} \, ,
\
\label{LTj}
\end{equation}
where  $\hat{P} = \partial_u(1-u^2)\partial_u, \hat{\Xi} = \partial_\xi(1+\xi^2)\partial_\xi$.
As we demonstrated in \cite{Pikina2022}, the right side of Eq. (\ref{LTj})  can be decomposed over the Legendre's polynomials in the form
 $\sum_n \sum_k \mathcal{W}_{n,k}\,\Xi_k\,P_n[u]$. Thus it is conveniently to find
 the auxiliary  response $T_{nm}$, which to the equation
\begin{equation}
[\hat{\Xi}+\hat{P}]\, T_{nm} = \frac{1}{4\xi_0^2} P_n[u] \Xi_m[\xi] \, ,\
\label{eqTnm}
\end{equation}
in this way one can obtain
\begin{equation}
    T_{nm}  = \frac{1}{4\xi_0^2}\frac{P_n[u]}{m(m+1)-n(n+1)}[\Xi_m[\xi] + W_{nm}\Xi_{n}[\xi]] \, , \
    \label{Tnm}
\end{equation}
where constant $W_{nm}$ is calculated from the boundary conditions of the
continuity of heat and the heat flux at the boundary of the drop, ($\xi>\xi_0$),  and using  condition of  damping of the critical temperature perturbation  in the air (outside the drop) at $\xi\to+\infty$ {(${T_{nm}(\xi>\xi_0)\propto \Xi_n^{(a)}}$)}, compare with \cite{Pikina2022}:
\begin{equation}
   W_{nm} = -\,\left\{\frac{\kappa\frac{\Xi_m}{\Xi_n^{(r)}} - \frac{\partial_\xi \Xi_m}{\partial_\xi \Xi_n^{(r)}}}{\kappa\frac{\Xi_n}{\Xi_n^{(r)}} - \frac{\partial_\xi \Xi_n}{\partial_\xi \Xi_n^{(r)}}}\right\}_{\xi=\xi_0}
\end{equation}
Gradually uncovering  the right part of the Eq. (\ref{LTj}) we found in the expression for critical response $T_{cj}$
\begin{multline}
    T_{c j} = 2\frac{T_{j+1,j-1} - T_{j-1,j+1}}{(2j-3)(2j+1)} -\\
    -(2j-1)\frac{T_{j+1,j-3} - T_{j-3,j+1}}{(2j-3)(2j+1)}
    -2\frac{T_{j-1,j-3} - T_{j-3,j-1}}{(2j-3)(2j+1)}  + \\
    c_{sj} \frac{T_{j-1,j+1} - T_{j+1,j-1}}{2j+1} +
    \frac{1}{c_{sj}}\frac{T_{j-3,j-1} - T_{j-1,j-3}}{2j+1} \, , \
    \label{Tcj}
\end{multline}
where $c_{sj} = - \mathcal{X}_{j-2}(\xi_0)/\mathcal{X}_{j}(\xi_0)$.

The obtained expression for $\psi_n$, Eq. (\ref{GenS}),  and Eqs. (\ref{solT1})--(\ref{Tcj}) allow us to find the general critical solution for the stream functions within the free drop and drop on the ring.

\section{Limitations of the critical Marangoni convection. The case  of a strongly oblate drop}
\label{sec:C}

Let us analyze the limit of a strongly flatten drop with $H/(2 R) \ll 1$  (i.e. the ellipticity ratio $\xi_0 \ll 1$).
Further we consider a small velocity deviation $\mathbf{v}_c$  from that of  the stationary velocity:
\begin{equation}
    \mathbf{v} = \mathbf{v}_{st} + \mathbf{v}_c \ .\
    \label{vcap}
\end{equation}
We are using the linearized Navier-Stokes equation and accordingly   Eq. (\ref{big1}) under condition
  that the nonlinear terms  $\mathbf{v}_{st}\nabla \mathbf{v}_c$ and $\mathbf{v}_{c}\nabla \mathbf{v}_{st}$ in Eq.(\ref{NSt01}) are  negligibly small in comparison with the viscous term:
\begin{itemize}
    \item $\rho \mathbf{v}_{st}\nabla \mathbf{v}_c$ and $\mathbf{v}_{c}\nabla \mathbf{v}_{st}$ are much smaller than $\eta \Delta \mathbf{v}_c$,  then $v_{st}\ll \eta/(\rho_0 H) \implies {\mathrm{Ma}\ll \frac{\eta}{\rho_0 \chi} (R/H)}$,
    \end{itemize}
    where we use  the definition of Ma, Eq. (\ref{Ma}),  and  estimations for the stationary velocity $v_{st}$ obtained in Sec. II C b: $v_{st} \sim (\varsigma/\eta)(H/R) T$. Keeping in mind that for the ordinary liquids ${\eta}/{(\rho_0 \chi)} \sim  10^3$ (see Appendix \ref{sec:A}), the above inequality is valid for the  values of Ma we exploit in Sec. III.

In similar way we omit the term  related to the heat transfer by a stationary
flux in the heat transport equation for the flatten drop:
\begin{itemize}
        \item $\mathbf{v}_{st}\nabla T_c$ is much smaller than $\mathbf{v}_c\nabla T_{st}$,   then $\qquad \qquad$ $v_{st}\ll \chi/H\,  \mathrm{Ma}_c \implies (R/H) \gg 1 $
\end{itemize}
where $T_{st}\approx T_0$.
Moreover, as we discussed in Sec. III in this case  the  velocity components
$\mathbf{v}_{st}$ and  $\mathbf{v}_c $ are essentially spatially separated   from  each other. Besides that the lateral  region  where   $\mathbf{v}_c $ varies essentially is of the order of $r_c\ll R$. This inequality results in  additional smallness proportional to $(H/R)^{- 1/2}$ which  further  justifies the removing of the nonlinear terms and  the term related to the stationary flux in Navier-Stokes and heat transport equations describing the critical solutions.

\section{Deviations of the drop shape from the oblate spheroid }
 \label{sec:D}

Here we analyze  the role othe small deviations of the drop surface from the shape of oblate spheroid.
We assume, that the drop shape   is  defined by an equation:
\begin{equation}
\Phi[\xi, u] = \xi  - \left(\xi_0 + \delta \xi[u]\right) = 0 \
    \label{fap}
\end{equation}
where $\delta \xi[u]$ represents a small deviation from the oblate spheroid form.
The normal vector to the surface is given by the canonical expression
$\bm n =\nabla \Phi/|\nabla\Phi|$.
Since the normal velocity component $v_n$ is equal to zero at a drop interface, the corresponding boundary condition is written in a form
\begin{equation}
    \psi[u, \xi[u]] = 0 \ . \
     \label{fap1}
\end{equation}
To resolve  the boundary condition (\ref{fap1}) the conventional perturbation theory  over $\psi[u, \xi]$ is used: $\psi[u, \xi] = \psi^{(0)}[u, \xi] + \psi^{(1)}[u, \xi]$. This leads to an equation
\begin{equation}
     \psi^{(1)}[u, \xi_0] + \delta\xi[u] \,\big\{\partial_\xi \psi^{(0)}[u, \xi]\big\}_{\xi=\xi_0}=0 \,. \
      \label{fap01}
\end{equation}
The distortion of an oblate drop shape results in the deviations of the temperature field from that obtained for the initial  drop: $T[u, \xi] = T^{(0)}[u, \xi] + T^{(1)}[u, \xi]$. Accordingly, the boundary condition for the temperature distribution takes the form
\begin{multline}
        T[u, \xi[u]] = T_{air}[u, \xi[u]] \to \\
       {T^{(1)} - T^{(1)}_{air}} + \delta\xi[u]\,{\partial_\xi \big(T^{(0)} - T^{(0)}_{air})\big)}=0 \ \text{ at } \xi=\xi_0  \ , \
     \label{fap2}
\end{multline}
In turn, the boundary condition far away from the drop surface reads
\begin{equation}
    T^{(1)}_{air}(\xi\to\infty) \, \to \, 0  \ . \
     \label{fap3}
\end{equation}
To write the condition of the equality of the heat fluxes at the drop interface we first need
  to find the normal vector $\mathbf{n}$ to the distorted  drop surface:
\begin{equation}
   \mathbf{n} = \mathbf{e}_\xi  - \mathbf{e}_u \, \frac{h_{\xi}}{h_u} \,\partial_u\delta \xi[u] \ . \
     \label{fap4}
\end{equation}
 Then  from the condition of the equality of the heat fluxes at the drop interface we obtain
\begin{widetext}
\begin{equation}
  {{\frac{\partial_\xi \big(T^{(1)}-\kappa T^{(1)}_{air}\big)}{h_\xi}}}  - (\partial_u \delta\xi[u])\, \frac{h_{\xi}}{h_u}\,\frac{\partial_u \big(T^{(0)}-\kappa T^{(0)}_{air}\big)}{h_u}
 + \delta\xi[u]\, {\partial_\xi{\frac{\partial_\xi \big(T^{(0)}-\kappa T^{(0)}_{air}\big)}{h_\xi}}} = 0 \text{ at } \xi=\xi_0  \ , \
     \label{fap5}
\end{equation}
which can be converted to the form
\begin{eqnarray}
  {{\partial_\xi \big(T^{(1)}-\kappa T^{(1)}_{air}\big)}}  - (\partial_u \delta\xi[u])\, \frac{(1-u^2)}{(\xi^{2} + 1)}\,{\partial_u \big(T^{(0)}-\kappa T^{(0)}_{air}\big)}} +\delta\xi[u] \, {\partial^2_\xi \big(T^{(0)}-\kappa T^{(0)}_{air}\big)
   \nonumber \\
   +\,\delta\xi[u]\, \left(\partial_\xi \big(T^{(0)}-\kappa T^{(0)}_{air}\big)\right)\, \frac{\xi \left(u^{2} - 1\right)}{\left(u^{2} + \xi^{2}\right) \left(\xi^{2} + 1\right)}= 0 \quad \text{ at } \xi=\xi_0  \ . \
     \label{fap6}
\end{eqnarray}

In turn,  Marangoni  boundary condition can be written as
\begin{equation}
    {\sigma^{(1)}_{u\xi}-\frac{\partial_u T^{(1)}}{h_u}} + \frac{h_{\xi}}{h_u}\,(\partial_u\delta\xi[u])\,(\sigma_{\xi\xi}^{(0)} - \sigma_{uu}^{(0)}-\partial_\xi T^{(0)}) + \delta\xi[u] \,\partial_\xi \Big(\sigma_{u\xi}^0 -  \frac{\partial_u T^{(0)}}{h_u}\Big)=0 \text{ at } \xi=\xi_0
     \ . \
     \label{fap7}
\end{equation}
\end{widetext}

Thus one can see from Eqs.  (\ref{fap01}), (\ref{fap2}), (\ref{fap6}), (\ref{fap7}),  that  for the case of the small smooth deviations
$\delta \xi[u]$ from the shape of oblate spheroid we have only small corrections to the results  for the  velocities and temperature distributions, obtained in  Secs. II and III.

\section{Сritical marangoni convection in a flat liquid layer}
\label{sec:E}

In Sec. III. C  we have used the formalism of the Stokes stream functions to study a crossover from an oblate drop to a flat liquid layer with two free interfaces. This was made by reducing of the droplet ellipticity ratio $\xi_0$ to zero, which leads in the limit
 $\xi_0\to 0, \kappa\to 0$ to the case of a flat film. Here we present the direct solution of Marangoni problem for a flat liquid layer using the conventional approach \cite{Koschmieder1974,Gershuni1972,Pikina2021,Pearson}. This allows us to compare the results for the critical Marangoni flow in a flat liquid film with that shown in Fig. \ref{freelay}, and to correct some results obtained in our previous paper \cite{Pikina2021}.

We remind that
 the system of  equations to describe  Marangoni convection in a flat liquid layer can be written as
 \begin{equation}
    \Delta \Delta v_{zc} = 0 \, , \ \Delta {T}_c = v_{zc} \, . \
    \label{lins}
\end{equation}
 In turn,  the boundary condition for the critical temperature distribution
in  the case of the flat liquid layer
should be written in the form (compare with \cite{Pikina2021})
     \begin{eqnarray}
     \frac{\partial {\widetilde{T}_c}}{\partial {\bf n}}\vert_{\widetilde{z}=0,1} = \,-\, b\,\widetilde{T}_c[\widetilde{z}=0,1] \ . \ \  \ \label{bc2}
      \end{eqnarray}
        where
     $b=a\,H/\varkappa\,$  is so-called Bio number -   a dimensionless coefficient of   convective heat transfer, and $a$ is the  coefficient of  convective heat transfer on the free surface of the drop  in the heat flux balance \cite{Koschmieder1974,Gershuni1972}.
     In terms of the conventional  boundary conditions
    of the equality of the
     temperature and  normal heat fluxes at the drop-air interface introduced in Sec. II, the expression for Bio number takes a simple form $b = k\kappa$.

     It is convenient to use below the cylindrical coordinates ($\widetilde{r}$, $\widetilde{z}$), where $\widetilde{z} = z/H$,  $\widetilde{r} = r/H$ are the  dimensionless coordinates in the direction of the film normal and in the lateral plane, respectively. The $\widetilde{z}$ values 0 and 1 correspond to the bottom and top film interfaces, respectively.
In such representation the decay of the critical perturbations of the velocity and temperature is described by  Bessel function of a zero order \cite{Pikina2021}:
      \begin{eqnarray}
  { \widetilde{ v}}_{cz} = V[\widetilde{z}] \,J_0[k \widetilde{r}],
    \nonumber \\
   {\widetilde{T}_c} = \Theta[\widetilde{z}] \, J_0[k \widetilde{r}] \ , \
    \label{cyl}
\end{eqnarray}
where $k$ is the  horizontal (in the lateral plane of a droplet) dimensionless wave vector.
We note that  Bessel function $J_0[k \widetilde{r}]$ appears in Eq. (\ref{cyl}) due to an axial symmetry of the problem.
 It is important that another possible type of cylindrical functions - Neyman function does not
appear in expansion (\ref{cyl})  as a consequence of the condition for regularity of the solution of Eq. (\ref{lins}) at  $r=0$.
The corresponding set of differential equations for the amplitudes
of the normal perturbations of the vertical velocity $ V[\widetilde{z}]$
and temperature $\Theta[\widetilde{z}]$ have a form  (see \cite{Pikina2021}):
\begin{align}
    k^{4} V - 2 k^{2} \frac{d^{2}}{d \widetilde{z}^{2}} V + \frac{d^{4}}{d \widetilde{z}^{4}} V = 0 \, , \
      \label{leq1}\\
    \left(k^{2} \Theta - \frac{d^{2}}{d \widetilde{z}^{2}} \Theta \right)  +  V = 0 \, , \
    \label{leq2}
\end{align}
with the boundary conditions for velocity: $V[\widetilde{z}=0]=0$, $V[\widetilde{z}=1]=0$.
The initial boundary conditions (\ref{bc2}) with account to the new variable, $\Theta[\widetilde{z}]$, take the form
 \begin{eqnarray}
          \frac{\partial { \Theta}}{\partial \widetilde{z}}\Big\vert_{\widetilde{z}=0} = \, b\, \Theta[\widetilde{z}=0] \ , \ \  \frac{\partial { \Theta}}{\partial \widetilde{z}} \Big\vert_{\widetilde{z}=1}  = -\, b\, \Theta[\widetilde{z}=1] \ , \
         \label{bcT2}
         \end{eqnarray}
        In turn, Marangoni  boundary conditions can be written as \cite{Pikina2021}:
    \begin{eqnarray}
   V^{''}_{zz} = \, \hbox{Ma} \,k^2\, \Theta \ \ ({\hbox{at}} \  \widetilde{z}=0)  \ , \
       \label{bc10} \\
         V^{''}_{zz} = \,-\,  \hbox{Ma}  \,k^2\, \Theta \   \  ({\hbox{at}} \  \widetilde{z}=1) \ . \
       \label{bc11}
      \end{eqnarray}

For $V[z]$ from  Eq. (\ref{leq1}) we have (likewise  as in \cite{Pikina2021}):
\begin{equation}
    V = C_{1} \underbrace{\left(\widetilde{z} - 1\right) \sinh{\left[k \widetilde{z} \right]}}_{V_1[\widetilde{z}]} + C_{2} \underbrace{\widetilde{z} \sinh{\left[k \left(1 - \widetilde{z}\right) \right]}}_{V_2[\widetilde{z}]} \, . \
     \label{Vz}
\end{equation}

\begin{figure}
    \centering
    \includegraphics[scale=0.7]{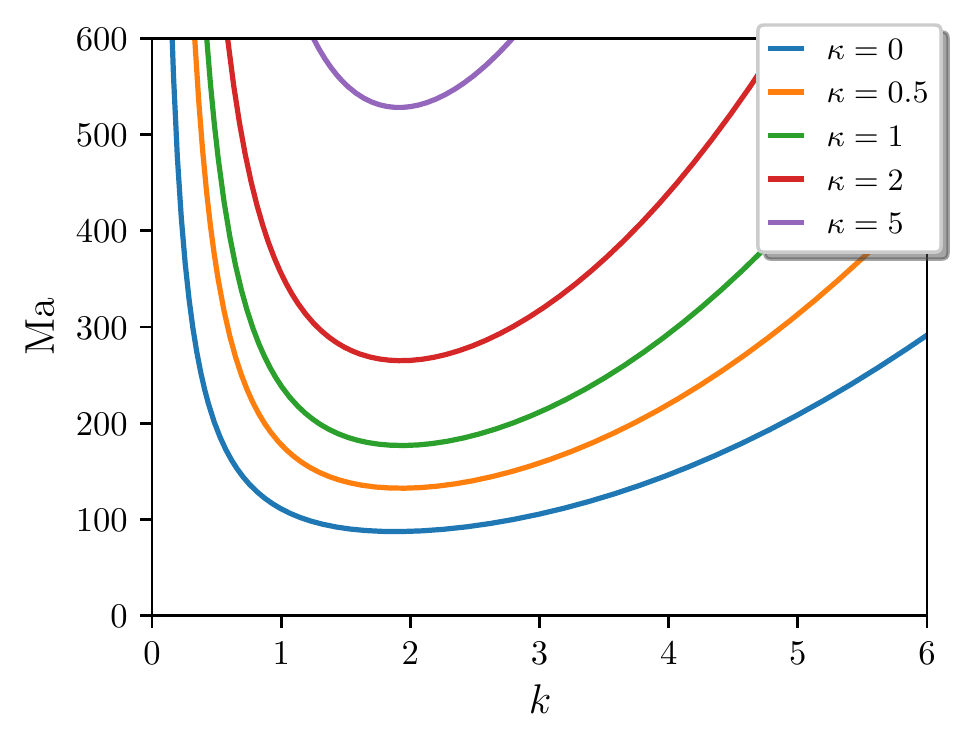}
    \caption{Dependencies of the Marangoni number, Ma, on the wave vector of in-plane instability $k$ for different values of the relative heat conductivity $\kappa$,  characterizing the magnitude of the heat transfer across the liquid layer.
    The critical wave number $k_c$ corresponds to the minimum $\mathrm{Ma_c}$ for the fixed $\kappa$.}
         \label{Figure1f}
\end{figure}

 The solution of Eq. (\ref{leq2}) can be found using the formalism of the Green functions  \cite{Pikina2021}. As a result the  Green function
  $G_{ \Theta}$  can be written as:
      \begin{eqnarray}
 \hskip-.0true cm
  G_{ \Theta}[k,\widetilde{z},\widetilde{z}'] = \frac{1}{W}\left\{\begin{aligned}
u[\widetilde{z}]\,v[\widetilde{z}'] \  , \  \widetilde{z} < \widetilde{z}' \  ,     \  \\
u[\widetilde{z}']\,v[\widetilde{z}] \  , \  \widetilde{z} > \widetilde{z}'  \ ,  \
 \end{aligned} \right. \
 \label{Gr2}
 \end{eqnarray}
 with
   \begin{eqnarray}
 u[\widetilde{z}]=
k\,\cosh[k \widetilde{z}] + b\,\sinh[k \widetilde{z}]  \ , \
 \label{uv1} \\
v[\widetilde{z}]=k\,\cosh[k (1 - \widetilde{z})] + b\,\sinh[k (1 - \widetilde{z})]  \ , \
 \label{uv2}
 \end{eqnarray}
 and Wronskian
\begin{equation}
  W = -\, k\, \big(2\, b\, k\, \cosh[k] + (b^2 + k^2)\,\sinh[k]\big)
  \ . \
 \label{Gr03}
 \end{equation}
Then, for  the critical temperature perturbation one obtains \cite{Pikina2021}
\begin{equation}
    \Theta = C_1 \Theta_1[\widetilde{z}]  + C_2 \Theta_2[\widetilde{z}]  \, , \
   \label{Thc1}
\end{equation}
where
\begin{equation}
    \Theta_{1,2}([\widetilde{z}] ) = \int_0^1 G[\widetilde{z} ,\widetilde{z}'] V_{1,2}[\widetilde{z}']  d{[\widetilde{z}] '}\, . \
    \label{Thc2}
\end{equation}

After substitution of Eqs. (\ref{Vz}), (\ref{Thc1}) to Eqs. (\ref{bc10}), (\ref{bc11}) we arrive to the system of equations:
 \begin{equation}
    \underbrace{\begin{pmatrix}
    V_1''(0) - \mathrm{Ma}\,k^2 \Theta_1[0] & V_2''[0] - \mathrm{Ma}\, k^2 \Theta_2[0]\\
    V_1''(1) + \mathrm{Ma}\,k^2 \Theta_1[1] & V_2''[1] + \mathrm{Ma}\,k^2 \Theta_2[1]
    \end{pmatrix}}_{\hat{M}} \begin{pmatrix}
        C_1\\
        C_2
    \end{pmatrix} = 0\, ,
    \label{Mal1}
\end{equation}
 \begin{figure}
    \includegraphics[scale=0.7]{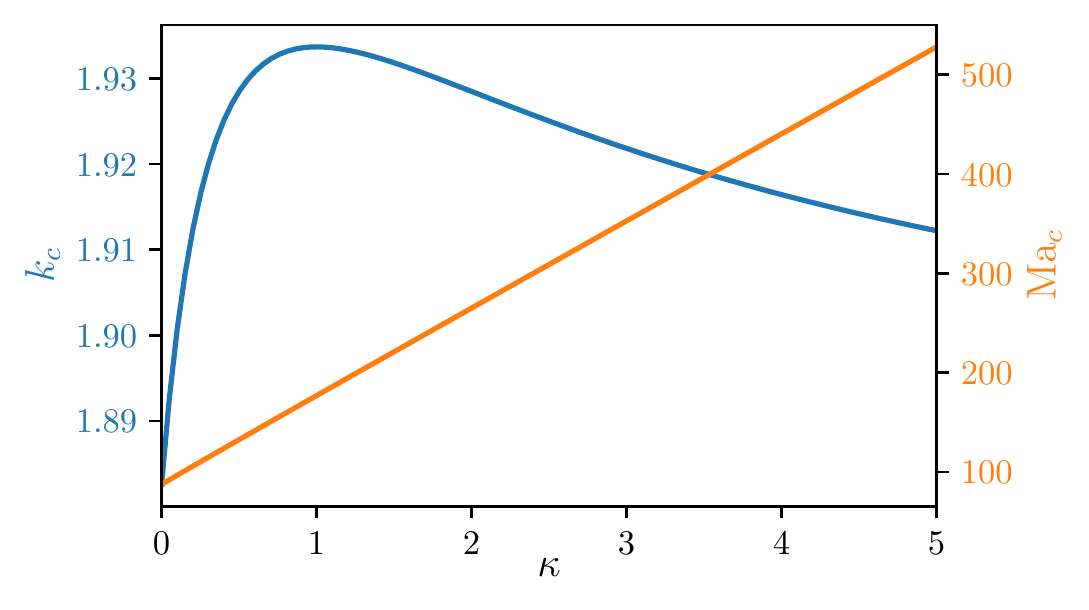}
    \caption{Dependencies of the critical  Marangoni number, Ma$_c$, and the  critical   dimensionless
 wave vector $k_c$ on the   relative heat conductivity $\kappa$. The $\mathrm{Ma}_c$  and $k_c$ correspond to the minimal values of the curves presented  in    Fig. \ref{Figure1f}.}
    \label{Figure2f}
\end{figure}
Finally, after solving the equation: $\det \hat{M}=0$, we derive an expression, determining   Marangoni number Ma$[k]$
as a function of the lateral  wave vector $k$ in dependence on the value of Bio number $b$:
   \begin{eqnarray}
 \hbox{Ma}^2[b,k]\,=\,
    \,256\, k^{2}\, \sinh^{2}[k]\,\times \,
   \qquad
    \nonumber\\     \times\,
   \left(b^{2} \sinh[k] + 2 b\, k \cosh[k] + k^{2} \sinh[k] \right)
    \nonumber\\
       \,\big\{8 k^3 \cosh[k] - (3 + 12 k^2 + 4 k^4) \sinh[k]
     \nonumber\\
     + \, \sinh[3 k]\big\}^{- 1} \ . \qquad
\label{Ma3}
 \end{eqnarray}
   The corresponding dependencies
of the Marangoni number, Ma, on the wave vector of in-plane instability $k$ for different values of the relative heat conductivity $\kappa$ are shown in Fig. \ref{Figure1f} (we use here the relationship $b = k\kappa$).
Let us emphasize the square dependence of Ma on the wave number $k$: $ \hbox{Ma}^2[b,k]$ in Eq. (\ref{Ma3}), which reflects the symmetry of a system relative to transformation $\mathrm{Ma}\to - \mathrm{Ma}$. This corresponds to a formal change of a sign (i.e. of the direction of the temperature gradient across the flat liquid layer).

The  dependecies of  Ma$_c$  and $k_c$ on relative heat conductivity $\kappa$ are shown in Fig.  \ref{Figure2f}. One can  see that in the limit $\kappa\to 0$  the parameters characterizing the formation of a set of critical vortices  in the lateral plane of a film constitute $\mathrm{Ma}_c\equiv \mathrm{Ma}_c^{(fl)} \approx 87.5$, $k_c \approx 1.9$. This is in accordance with our results on the critical Marangoni instability in a flat liquid layer obtained in Sec. III. C, see Fig. \ref{freelay}.

\section{Сritical Marangoni convection in a liquid drop with a slowly varying height}
\label{sec:F}

In the previous Appendix we have solved the problem of the critical Marangoni convection in an infinite flat layer. These results can be used to solve the similar problem in a strongly oblate liquid drop, which is characterized  by a slowly decaying height and can be considered as a  sequence of the locally flat layers.
 The drop height $h[r]$ slowly diminishes upon an increase of a lateral distance $r$ from the rotational axis of the drop of an axial symmetry.
 At the same time, a  constant temperature gradient  $A$ is applied accross the flattened drop.
 All coordinates are measured in the units of the height $H$ in the center of the drop, the temperature is given in the units of $A\,H$  and velocity in  $\chi/H$, i.e. $r = H \tilde{r}, \tilde{r} \in (0, 1/\epsilon)$,
$z = H \tilde{z}, h = H \tilde{h}, \tilde{h} = \sqrt{1 - \tilde{r}^2 \epsilon^2}$, where $\epsilon = H/R\ll\,1$. In further derivations we use the dimensionless quantities, and omit the tilde symbol for simplicity. The system of  equations describing the thermocapillary convection in a locally flat liquid layer is given by  Eq. (\ref{lins}). Let us define  the positions of  the bottom and upper free surfaces of the locally flat layer as   $z=\pm h/2$.
 Since the first derivative of a drop current height $h'[r]=dh/dr$ possesses the higher order of smallness  with respect to $h$ ($h'\sim r \epsilon^2$, and on the localization scale $r\sim 1/\sqrt{\epsilon}$ one has  $h'\sim \epsilon ^{3/2}\ll\,1$)
 we can introduce the new vertical coordinates \u{z}$ =(z + 1/2)h$. In such a way
 the positions of the boundary free surfaces of a locally flat layer are  shifted to \u{z}$=0,1$, likewise it was made  in Appendix \ref{sec:E}.
 In that follows we omit the sign \u{\,} for the sake of simplicity. The boundary conditions at the free surfaces of the locally flat layers in this case are given by Eq. (\ref{bc2}) (see Appendix \ref{sec:E}).
  Let us rewrite the system of  equations to describe the Marangoni convection in a drop with a slowly varying height:
   \begin{eqnarray}
    (\Delta_\perp + \frac{1}{h^2}\partial_{z }^2 )^2 v_z = 0 \, , \
    \label{lins1} \\
   (\Delta_\perp + \frac{1}{h^2}\partial_{z }^2 )T_c = v_z \, , \
    \label{lins2}
\end{eqnarray}
   where index "$\perp\,$" in Laplacian relates  to the space derivatives in the lateral drop direction. While deriving Eqs. (F.1) and (F.2) the first derivative of a drop current height $h'[r]=dh/dr$ has been omitted due to the higher order of smallness (see above).

At the next step let us  expand the critical velocity $v_{z}$ and temperature deviation $T_c$ over Bessel functions of the zero order
\begin{eqnarray}
    {v_{z} = \int_0^\infty k\,d\,{k} V_k[z]\, J_0[k r/h[r]]}\, , \
\label{lins3} \\
    T_c = \int_0^\infty k\,d\,{k} \Theta_k[z]\,{h^2}[r]\, J_0[k r/h[r]] \, . \
    \label{lins4}
\end{eqnarray}
Thus, the problem under consideration is reduced to the case of the flat layer of a current height $h[r]$ (see Eqs. (\ref{Vz}) and (\ref{Thc1}) in Appendix E)
\begin{eqnarray}
    V_k(z ) = c_1[k]V_{1k}[z] + c_2[k]V_{2k}[z]\, , \
\label{lins5} \\
    \Theta_k(z ) = c_1[k]\Theta_{1k}[z] + c_2[k]\Theta_{2k}[z] \, , \
\label{lins6}
\end{eqnarray}
where
 $c_{1,2}[k]$ are the corresponding amplitudes.
After
substitution  of the expansions for  $v_{z}$ and $T_c$ (Eqs. (\ref{lins3}) and (\ref{lins4})) to   Marangoni boundary conditions
\begin{align}
    \partial_{z }^2 v_z= -{\mathrm{Ma}\,h^2}\, \Delta_\perp T_c \;(\text{at } \tilde{z}=0)  \, , \
    \label{lins7} \\
    \partial_{z }^2 v_z= {\mathrm{Ma}\,h^2}\, \Delta_\perp T_c \;(\text{at } \tilde{z}=1)  \, , \
    \label{lins8}
\end{align}
we obtain  the following integral equations
\begin{align}
    \int_0^\infty k\,d\,{k} \big(V_k''[0] - \mathrm{Ma}\, k^2 \,h^2[r]\,\Theta_k[0]\big)\,J_0[kr/h] = 0 \, , \,
    \label{lins9} \\
    \int_0^\infty k\,d\,{k} \big(V_k''[1] + \mathrm{Ma}\, k^2 \,h^2[r]\,\Theta_k[1]\big)\,J_0[kr/h] = 0 \, . \,
  \label{lins10}
\end{align}

Let us introduce the corresponding matrices  to represent the integral equations (\ref{lins9}), (\ref{lins10}) in  a more compact form:
\begin{equation}
    \hat{\mathcal{V}}_k = \begin{pmatrix}
        V''_{1k}[0] & V_{2k}''[0]\\
        V''_{1k}[1]) & V_{2k}''[1]
    \end{pmatrix}, \;
    \hat{\mathcal{T}}_k = k^2\begin{pmatrix}
        \Theta_{1k}[0] & \Theta_{2k}[0]\\
        -\Theta_{1k}[1] & -\Theta_{2k}[1] \,
    \end{pmatrix}
   \label{matr}
\end{equation}
It is also convenient to write the amplitudes of the solutions (\ref{lins5}) and (\ref{lins6}) as a two-component vector $\mathbf{c}[k]$
\begin{equation}
    \mathbf{c}[k] =
    \begin{pmatrix}
        c_1[k]\\
        c_2[k]
    \end{pmatrix} \, . \
    \label{vect}
\end{equation}
In such a way the system of equations (\ref{lins9}), (\ref{lins10}) can be rewritten as
\begin{equation}
    \int_0^\infty k dk J_0[kr/h] \left(\hat{\mathcal{V}}_k \mathbf{c}[k]- \mathrm{Ma}\,h^2[r]\,\hat{\mathcal{T}}_k \mathbf{c}[k]\right)=0 \ . \
    \label{systa1}
\end{equation}
 Let us apply the above formalism to  find the solution for the critical Marangoni flow in the
      flat liquid layer,  Appendix \ref{sec:E}. In this case for each wave vector  $k$ of the lateral perturbations there are two
 vectors $\mathbf{c}_\pm[k]$
\begin{equation}
    \hat{\mathcal{V}}_k \mathbf{c}_\pm[k] = \pm \mathrm{Ma}^{(fl)}[k] \hat{\mathcal{T}}_k\mathbf{c}_\pm[k] \, , \
     \label{systa2}
\end{equation}
which are the solutions of the system of equations (\ref{Mal1}) combined with the corresponding solution for  Marangoni number Ma$[k]$, determined by Eq.   (\ref{Ma3}).

Then, it is convenient to find the solution $\mathbf{c}[k]$  of the system of equations (\ref{matr}) on the basis of vectors $\mathbf{c}_\pm[k]$
 with  the corresponding amplitudes $a_+[k], a_-[k]$ in the form
\begin{eqnarray}
    \mathbf{c}[k] = a_+[k]\mathbf{c}_+[k] + a_-[k]\mathbf{c}_-[k] \, , \
     \label{vectc} \\
 \hbox{where} \qquad    \mathbf{a}^T = \begin{pmatrix}
        a_+[k] & a_-[k]
    \end{pmatrix}
  \, . \,   \label{vecta}
\end{eqnarray}

It is important  that we deal  with Ma  numbers in the vicinity of the critical $\mathrm{Ma}_c$ for the strongly oblate drop,
i.e. with Ma complying to  inequality  $\mathrm{Ma} - \mathrm{Ma}_c^{(fl)} \ll \mathrm{Ma}_c^{(fl)}$, where $\mathrm{Ma}_c^{(fl)}$ is a critical Marangoni number for an infinite flat layer considered in Appendix \ref{sec:E}.
  It means that the range of admissible $k$-values in the  solutions for $\mathbf{a}[k]$  is  localized  in the narrow vicinity of the optimal  wave vector $k_c$ corresponding to the minimum  of the
 upper branch of  $\mathrm{Ma}^{(fl)}[k]$ spectrum, determined by Eq.   (\ref{Ma3}) (compare with Fig. \ref{Figure1f}  where  Ma$[k]$ curves are extremely flat in the vicinity of the critical wave number $k_c$).
In this case one can expand $\mathrm{Ma}^{(fl)}[k]$ function near $k_c$ as
\begin{equation}
\mathrm{Ma}^{(fl)}[k]\approx \mathrm{Ma}_c^{(fl)} + \alpha_0\frac{(k-k_c)^2}{2} \, , \
\label{makfl}
\end{equation}
  where
$\alpha_0 =  \frac{d^2 \mathrm{Ma}^{(fl)}}{dk^2}\Big\vert_{k=k_c}$.

 After substitution of Eq. (\ref{vectc})   and an expansion (\ref{makfl}) to Eq. (\ref{systa1}) the later can be rewritten in the form
\begin{equation}
     \int_0^\infty k dk  J_0[k\rho] \left(\mathrm{Ma}^{(fl)}[k]- \mathrm{Ma}\,h^2[r]\,\right) { a_+[k]} \hat{\mathcal{T}}_k \mathbf{c}_+[k]\approx 0 \, , \
     \label{systa3}
\end{equation}
where $\rho = r/h[r]$. In accordance with the expression for the current drop height $h[r]$ one obtains: $r = \rho \sqrt{\frac{1}{\epsilon^{2} \rho^{2} + 1}}$,  $h^2 = 1 - \epsilon^2 r^2$, that leads to  $h^2 = \frac{1}{\epsilon^2\rho^2+1}$.
 According to Eq. (\ref{frc}),   the critical Marangoni flow is  localized
 within the "allowed" region  of a lateral size $r\sim {1}/{\sqrt{\epsilon}}$, thus  the square of  the current layer height can be expanded as  $h^2 \approx 1  - \epsilon^2 \rho^2$.
Then, we can use Hankel transformation over variable $k$ and convert the term  proportional to $\rho^2$ in the integral
representation (\ref{systa3}) of corresponding Hankel transform.
While doing that we use an advantage of the property of Hankel transformation:
$\int_0^\infty k\,dk\,J_0[k\rho]\,(- \rho^2)\,f[k]= \int_0^\infty k\,dk\,J_0[k\rho]\,(\partial_k^2 + 1/k\partial_k)\,f[k]$.
 Since the functions $J_0[k\rho]$ are linearly independent for different $k$, to solve  Eq. (\ref{systa3})  the
 integrand in  equation (\ref{systa3})  should be equated with  zero for any $k$:
\begin{equation}
     \left(\mathrm{Ma}^{(fl)}[k]-\mathrm{Ma}   - \mathrm{Ma}\epsilon^2(\partial_k^2 + 1/k\partial_k)\right) { a_+[k]} \hat{\mathcal{T}}_k \mathbf{c}_+[k]\approx 0 \, .\,
     \label{systa4}
\end{equation}

Therefore, the non zero values of the amplitudes $a_+[k]$ are concentrated in the narrow range   $\delta k\ll 1$,  whereas    $\hat{\mathcal{T}_k}\mathbf{c}_+[k]$ varies at a scale of an order of unity. It means that the main contribution from the differentiation  in Eq. (\ref{systa4}) results from the term $a_+[k]$. Moreover, due the same reason one can neglect the contribution from the operator $1/k\partial_k \sim 1/\delta k$ comparatively to the contribution from   $\partial_k^2 \sim 1/\delta k ^2$.
In doing so we arrive  from   Eq. (\ref{systa4}) to a simplified  equation
\begin{equation}
    \left(\mathrm{Ma}^{(fl)}[k]-\mathrm{Ma}   - \mathrm{Ma}\epsilon^2\partial_k^2 \right) { a_+[k]} = 0\,  . \
     \label{systa5}
\end{equation}
After  substitution of the quadratic expansion for $\mathrm{Ma}^{(fl)}$ in the  vicinity of $k_c$, Eq. (\ref{makfl}),  into Eq. (\ref{systa5}) one  obtains the well known equation for the harmonic oscillator \cite{Landau3}  in the main order over $\epsilon$:
\begin{equation}
    \left[- \mathrm{Ma}_c^{(fl)}\epsilon^2\partial_k^2 + \alpha_0\frac{(k-k_c)^2}{2}\right]{a_+[k]} = \delta \mathrm{Ma} \;{a_+[k]}\,  , \
     \label{systa6}
\end{equation}
where $\delta \mathrm{Ma} = \mathrm{Ma} - \mathrm{Ma}_c^{(fl)}$.
At this point it is convenient to introduce some notional
mass $m = \frac{1}{2 \mathrm{Ma}_c^{(fl)} \epsilon^2}$ and corresponding frequency ${\omega^2 = \alpha_0/m}$
 \cite{Landau3}.
 Using the known result from \cite{Landau3} one can write  the  exact expression for the discrete spectrum
of the  harmonic oscillator:

\noindent
\begin{equation}
\delta \mathrm{Ma}_n = \omega (n+1/2) = \epsilon \sqrt{2\alpha_0 \mathrm{Ma}_c^{(fl)}}(n+1/2) \, . \
 \label{dMa1} \end{equation}
   We note that our initial assumption about localization of the critical mode is fulfilled
   under  the following condition
\begin{equation}
   \delta k \sim  1/\sqrt{\omega m} = \sqrt{\epsilon}\left(\frac{2\mathrm{Ma}_c^{(fl)}}{\alpha_0}\right)^{1/4}\ll 1\, , \
\end{equation}
 i.e.  for $\epsilon \ll 1$.
We also emphasize that the above relations are valid  for $\delta \mathrm{Ma} \ll \mathrm{Ma}_c^{(fl)}$, i.e.  for the low-lying levels of
 $\delta \mathrm{Ma}$, see Fig. \ref{Figure1f}. From that   one more necessary condition arrives:
$\omega n \ll \mathrm{Ma}_c^{(fl)}\implies n\ll 1/{\epsilon}$.
Finally, we obtain an  expression  for  $\delta \mathrm{Ma}_n$ spectrum
\begin{equation}
    {\delta \mathrm{Ma}_n = \frac{H}{R} \,\sqrt{\alpha_0 \mathrm{Ma}_c^{(fl)}}(n+1/2),\; n\in\mathbb{N}} \, . \
     \label{systan}
\end{equation}

Therefore, the expression for the critical $\delta \mathrm{Ma}$ value ($n=0$) can be written as:
\begin{equation}
   \delta \mathrm{Ma}_c = \frac{H}{2R} \, \sqrt{\alpha_0 \mathrm{Ma}_c^{(fl)}} \, . \
   \label{dMac}
\end{equation}
The Eq. (\ref{dMac}) is used to derive an expression for the lateral localization of the critical convection mode in the strongly oblate drop, see Eqs. (\ref{mdMac}), (\ref{frc}) in Sec. III B.


\begin{thebibliography}{46}
\bibitem{psi} The stream function $\psi\,$  determines the instant fluid flow rate divided by $2\,\pi\,$ (the half of the total spatial angle) \cite{Happel}. To represent  $\psi\,$ in dimensionless form it is scaled by $\chi c^2/H$, and as such is used consitently throughout the paper.
\end{thebibliography}
\end{document}